\documentclass{amsart}
\usepackage[dvips]{epsfig}
\usepackage{graphicx}
\usepackage{latexsym}
\usepackage{amsmath}
\usepackage{amsthm}
\usepackage{amssymb}
\usepackage{booktabs}

\RequirePackage[colorlinks=true,citecolor=blue,urlcolor=blue]{hyperref}
\usepackage{scrextend}
\usepackage{ulem}
\usepackage{xcolor}
\usepackage{caption}
\usepackage{subcaption}
\usepackage{float}
\usepackage{hyperref}
\usepackage{comment}
\usepackage[applemac]{inputenc}
\usepackage{eurosym}
\newtheorem{corollary}{Corollary}
\newtheorem{definition}{Definition}

\newtheorem{proposition}{Proposition}
\newtheorem{remark}{Remark}

\newcommand{\vip}{\vskip.2cm}
\newcommand{\R}{{\mathbb{R}}}

\newcommand{\E}{\mathbb{E}}
\newcommand{\COMMENTAIRE}[1]{}

\begin{document}

\title[]{A Common Shock Model for multidimensional electricity intraday price modelling with application to battery valuation}

\author{Thomas Deschatre \and Xavier Warin}

\address{Thomas Deschatre, EDF Lab Paris-Saclay and FiMe, Laboratoire de Finance des March\'es de l'Energie, 91120 Palaiseau, France}
\email{thomas-t.deschatre@edf.fr}
\address{Xavier Warin, EDF Lab Paris-Saclay and FiMe, Laboratoire de Finance des March\'es de l'Energie, 91120 Palaiseau, France}
\email{xavier.warin@edf.fr}

\begin{abstract} 
In this paper, we propose a multidimensional statistical model of intraday electricity prices at the scale of the trading session, which allows all products to be simulated simultaneously. This model, based on Poisson measures and inspired by the Common Shock Poisson Model, reproduces the Samuelson effect (intensity and volatility increases as time to maturity decreases). It also reproduces the price correlation structure, highlighted here in the data, which decreases as two maturities move apart. This model has only three parameters that can be estimated using a moment method that we propose here. We demonstrate the usefulness of the model on a case of storage valuation by dynamic programming over a trading session. 

\end{abstract}

\maketitle

\textbf{Mathematics Subject Classification (2020)}: 60G55; 60G57; 62P05; 91G30.

\textbf{Keywords}: Electricity intraday prices, Poisson measures, Dependence, Storage valuation, Dynamic programming, Common Poisson Shock Model.

\section{Introduction}

\subsection{Motivation} The strong growth in renewable energy capacity in Europe is leading to increased liquidity in the short-term markets, particularly in the intraday electricity market. This market follows on from the spot market, which is an auction market where producers and suppliers submit forecasts of supply and demand curves for the next 24 hours. The spot market, also known as the day-ahead market, takes place at noon on the day before the delivery period and the intraday market starts at 3 pm. The intraday market is an order book market where transactions can take place continuously. It is possible to place bid or ask orders for any product up to at least one hour before delivery (e.g. up to 5 minutes before delivery in Germany). This allows producers and suppliers to adjust their positions in relation to their forecasts at the time of the spot market. For example, if a wind producer has bid too high on the spot market compared to his/her new forecast and that bid has been accepted, he/she can buy on the intraday market for that hour.  It should be noted that there are 24 delivery hours and therefore 24 markets operating simultaneously; however, {\it the intraday market} is used to refer to these 24 products. For more information on how the intraday electricity market works, see \cite{balardy2022}.

\smallskip
With the development of renewable energy, storage devices such as batteries are also being developed, allowing, for example, a renewable energy producer to store or release electricity to balance his/her commitment on the spot market, rather than buying or selling on the intraday market. More generally, a market participant, whether a producer or a supplier, may acquire a means of storage to facilitate the balancing of supply and demand, but also to capture a value associated with market volatility. It is therefore important to quantify the value that can be extracted from the market through storage. In addition, the battery makes it possible to link different maturities: without storage, it is not possible for an actor to buy electricity for one delivery time and sell it at another delivery time without incurring a very large financial penalty. The value of the battery then depends not only on the price level, but also on the level of volatility and correlation. Stochastic optimisation has been used for decades by energy producers to manage water and gas reservoirs using the principle of dynamic programming but with low-dimensional models for prices, see \cite[Section 3.3]{machlev2020}. The first methods were tree methods, and more recently the Longstaff-Schwarz algorithm \cite{longstaff2001valuing} has been adapted to this problem \cite{warin2012gas}. More recently, these methods have been adapted to optimise batteries taking into account their yield (see the review of Machlev et al.~\cite{machlev2020} on this topic).\\
Optimising and valuing a storage at the trading session level requires a multidimensional pricing model that captures the correlation structure of prices: in \cite{finnah2022integrated}, Finnah et al. optimise a battery on both the day-ahead and intraday markets using a simple high-dimensional autoregressive model for price forecasts. Because of the very high dimensionality of the price model, they use an approximate dynamic programming to correctly optimise the asset and show that taking into account the high dimensionality of the prices is necessary to get good results.

\smallskip
To the best of our knowledge, the literature on stochastic modelling of intraday prices at the session level is very scarce. Both Favetto \cite{favetto2019} and Graf von Luckner and Kiesel \cite{graf2020} are interested in modelling the arrival of order books using Hawkes processes. Blasberg et al.~\cite{blasberg2019} follow the modelling of Graf von Luckner and Kiesel \cite{graf2020} and are interested in the evolution of the parameters over time, while Kramer and Kiesel \cite{kramer2021} integrate covariates into the model (errors in the forecast of renewable generation and the activated volume on the balancing market) on which the intensity depends. This modelling is unidimensional, i.e. it only treats maturities independently. In addition, the modelling of order arrivals is not sufficient for the valuation of storage assets. Deschatre and Gruet~\cite{deschatre2023b} are interested in the modelling of prices and thus add the marked aspect to the modelling of the order book: a mark represents the price return associated with an event. The use of a marked bivariate Hawkes process with an exponential baseline intensity provides a good representation of prices. The baseline intensity makes it possible to represent the Samuelson effect, while the self-excitation makes it possible to represent the microstructure noise and in particular the signature plot, which is classical in finance \cite{bacry2013a,bacry2013b}. Once again, the price modelling is only one-dimensional and does not take into account the dependence between the different maturities, which is essential for the valuation of assets with payoffs that depend on several maturities. Narajewski and Ziel~\cite{narajewski2020} and Hirsch and Ziel~\cite{hirsch2022, hirsch2023} develop simulation-based forecast models which are similar to econometric models \cite{kiesel2017, kremer2021}, that look at the relationship between intraday price level and fundamental variables such as changes in solar and wind generation forecasts and changes in demand forecasts, but which look at the distribution of returns (variance, skewness, kurtosis) in addition to the price level. The model of Hirsch and Ziel in \cite{hirsch2023} is a multidimensional version of their model in \cite{hirsch2022} and is to our knowledge the first model to focus on the dependence structure of prices in a simulation framework.


\subsection{Main results} In this paper, we propose a pure simulation model of multidimensional intraday electricity prices at the level of the trading session, allowing the valuation of assets such as storage facilities. First, we carry out an empirical analysis of transaction prices on the French and German intraday markets and highlight two particularly important stylised facts that we wish to represent: 
\begin{itemize}
\item[(i)] the Samuelson effect, corresponding to the increase in the intensity and volatility of price movements over time during the trading session and already identified and modelled by Deschatre and Gruet~\cite{deschatre2023b} ;
\item[(ii)] the forward structure of the price correlation matrix, which is the main novelty: we observe a decrease in correlation with the distance between several maturities; this relationship is known on commodity futures markets and can be called the Samuelson correlation effect \cite{schneider2018}.
\end{itemize}
These empirical results constitute the Section~\ref{sec:empirical_facts}.

\smallskip
Next, we propose a multidimensional price simulation model that allows the simultaneous representation of (i) and (ii). Due to the nature of the data, which correspond to asynchronous transactions, we choose a pure jump model with jump times corresponding to transaction times and jump sizes corresponding to the sizes of the returns. The model, described in Section~\ref{sec:model}, is constructed using three-dimensional Poisson measures. Marginally, each price is a compound inhomogeneous Poisson process of intensity increasing with time, as in \cite{deschatre2023b} if we neglect the stochastic part of the intensity. A marked compound inhomogeneous Poisson process can be constructed from a two-dimensional measure whose first dimension corresponds to the time aspect and the second to the jump law. The third dimension in our model makes it possible to introduce dependency between products of different maturities. The proposed model is strongly inspired by the Common Shock Poisson Model \cite{powojowski2002, lindskog2003} and adapted to electricity markets. There are two types of price movements: some are specific to each maturity and others will affect the different maturities simultaneously, in the form of a common shock that may, for example, represent the failure of an asset affecting several hours of delivery. To represent the Samuelson correlation effect, the shocks can only affect successive maturities, starting with the nearest, and the probability of affecting the next $m$ successive maturities decreases with $m$. The model is parameterised by only 3 parameters in addition to the law of jumps, and each parameter has a simple interpretation: 
\begin{itemize}
\item[-] the first one is the rate of increase in volatility as a function of time and the rate of decrease in correlation as a function of the distance to maturity ;
\item[-] the second corresponds to the intensity of transactions taking place independently on each maturity ;
\item[-] the third one is the intensity of transactions that can take place simultaneously on several maturities and which result from a common shock.
\end{itemize}
Unlike \cite{deschatre2023b}, our model does not allow microstructure noise to be represented, although it is taken into account in the estimation procedure. Since our objective is to optimise and make the most of a storage  asset, which we are not looking at at very high frequency, we have neglected this aspect. A very simple based moments estimation procedure is associated with the model and is given in Section~\ref{sec:estimation}. Although there are similarities with the simulation-based forecast model of Hirsch and Ziel~\cite{hirsch2023}, our approach is quite different. The aim is to have a simulation model of price trajectories, the objective of which is not to make forecasts as in \cite{hirsch2023} but to value assets by stochastic optimisation. In a pure simulation context, we would need a simulation model for each covariate used for forecasting if we wanted to simulate price trajectories every day with the model of Hirsch and Ziel~\cite{hirsch2023}, which seems complicated in practice. Our model makes it very easy to simulate multidimensional price trajectories for a given day, with a very limited number of parameters compared to a forecasting model. On the other hand, our model is not suitable for ensemble price forecasting. An interesting intermediate between our model and the one of Hirsch and Ziel~\cite{hirsch2023} would be to add a limited number of covariates that we can simulate to our model to explain the intensity of price changes, as Kramer and Kiesel \cite{kramer2021} do for order books. To our knowledge, our model is therefore the first pure simulation multidimensional model of intraday electricity prices at the scale of a trading session. There are models for intraday prices, such as in~\cite{finnah2022integrated}, but at a higher time scale, e.g. representing only the last transaction price or an index at daily granularity, but not the continuous evolution of transaction prices during the trading session.

\smallskip
Finally, we propose a method for valuing a storage device in Section~\ref{sec:battery}, in this case a battery, by dynamic programming using our simulation model. The optimisation method is then back-tested on the data using the controls learned from the simulation model, and provides battery values for the French and German markets over the period from 2019 to 2022. The values obtained show the relevance of the chosen statistical model and show that the model exhibits strategies far more interesting than  deterministic strategies derived from the day-ahead prices.

\section{Empirical stylised facts}
\label{sec:empirical_facts}

In this section, we highlight some of the stylised facts that led us to choose the model described in Section~\ref{sec:model}. The two main ones are (i) the structure of intensity as a function of time to maturity, and therefore of volatility, also known as the Samuelson effect, already identified by Deschatre and Gruet~\cite{deschatre2023b}, and since we focus on multidimensional price modelling, (ii) the very specific correlation structure of the electricity intraday market highlighted for the first time in this paper but already present in \cite{hirsch2023}. We also recall the presence of microstructure noise in our data, which is important to take into account for the estimation, but which we will not model, unlike Deschatre and Gruet~\cite{deschatre2023b}, which do so in a one-dimensional case. 

\subsection{Dataset} 
\begin{figure}[H]
\begin{tabular}{cc}
  \includegraphics[width=0.45\textwidth]{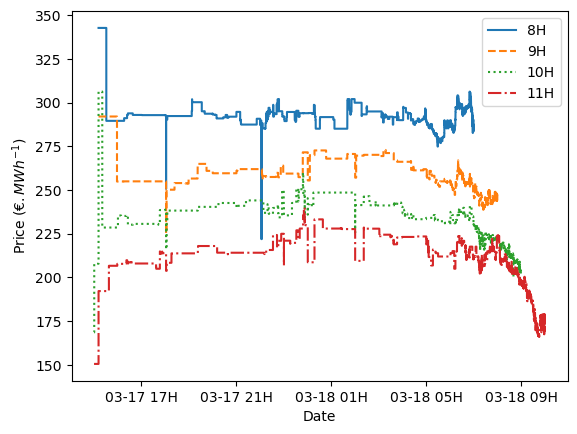}   &  \includegraphics[width=0.45\textwidth]{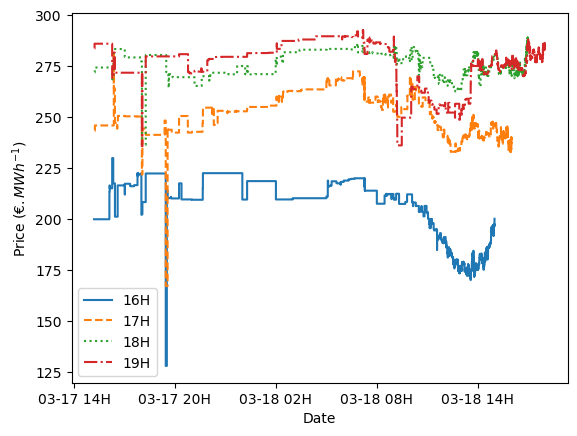} \\
\end{tabular}
    \centering
    \caption{German transaction prices for the trading session on March 18, 2022.}
    \label{fig:price_ts}
\end{figure}

The dataset provided by EPEX Spot consists of the German and French electricity intraday transaction prices for products with a delivery period of one hour between December 1\textsuperscript{st}, 2018 and December 31\textsuperscript{st}, 2022 for Germany and France. For each day $d$, there are 24 products corresponding to the 24 hours of the day. These 24 products correspond to 24 markets and for each market, the trading session starts at 3 pm on the day before the delivery, i.e. $d-1$, and ends between thirty and five minutes before the start of delivery, depending on the year and the country. These markets are order book markets and the prices evolve continuously throughout the trading session. Transaction prices consist of timestamps accurate to one minute (2019), one second (2020) or one millisecond (2021 and 2022), and associated prices corresponding to a transaction taking place at that time. If more than one price has the same timestamp, only the last one is taken into account. The tick size is 0.01\euro{}/MWh. As one hour before delivery, it is no longer possible to trade across countries, and thirty minutes before delivery, it is only possible to trade at a smaller zone level within the country, we only consider data up to one hour before delivery as in \cite{deschatre2023b}. For illustrative purposes, German transaction prices are shown in Figure \ref{fig:price_ts} for some maturities on March 18, 2022.
Very large returns, sometimes greater than 100\euro{}/MWh (i.e. 10000 ticks), are identified and can affect the different estimated quantities. These large returns, which often occur at the beginning of the trading session, are often followed by a return of the same magnitude in the opposite direction. This seems to be analogous to bouncebacks mentioned in \cite[Chapter 2, Section 4.2]{jacod2013}. For each delivery period in the following, we remove the returns that are greater than 5 times the standard deviation of all the returns. The number of removed returns is less than $1\%$ for each period. Although only presented for Germany, all the empirical results presented below also hold for France.

\subsection{Increasing intensity of price changes} 
\begin{figure}[H]
\begin{tabular}{cc}
  \includegraphics[width=0.45\textwidth]{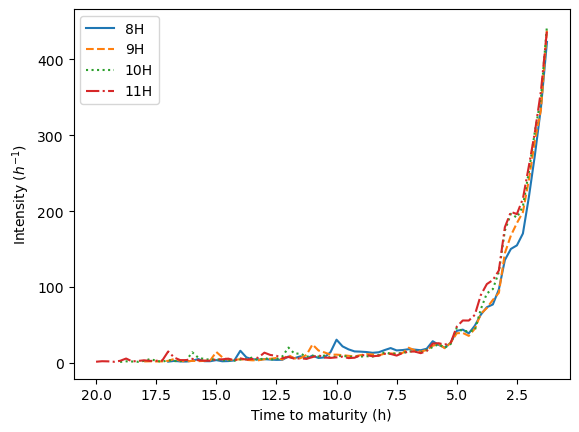}   &  \includegraphics[width=0.45\textwidth]{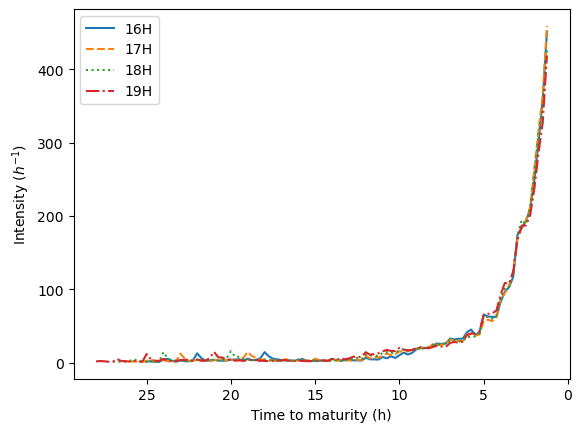} \\
\end{tabular}
    \centering
    \caption{Intensity of German transaction price changes in 2022 for some delivery periods.}
    \label{fig:intensity}
\end{figure}

Favetto \cite{favetto2019} and Graf von Luckner and Kiesel \cite{graf2020} identify an exponential increase in order book activity with trading time. Deschatre and Gruet~\cite{deschatre2023b} also identify this phenomenon for the intensity of mid-price changes and considers a model where the non-stochastic part of the intensity of mid-price moves is an exponential function of time. This phenomenon is particularly important as it has direct implications for volatility modelling: in \cite{deschatre2023b}, increasing intensity induces an exponential increase in volatility as time to maturity decreases, corresponding to the so-called Samuelson effect, which is consistent with the data. Samuelson's effect is studied by A\"id et al.~\cite{aid2022} for intraday markets; they develop an equilibrium price model and show conditions under which this effect holds. Samuelson's effect is also very common in commodity futures markets and especially in electricity futures markets, see \cite{jaeck2016}. Intensity curves for different maturities are shown in Figure~\ref{fig:intensity} for German transaction prices in 2022. For a given maturity, the corresponding intensity curve has been calculated from the average number of price changes over the different trading sessions of the year in 15-minute windows. This exponential increase is observed for the other maturities and for the rest of the data set, for Germany but also for France.

\subsection{Microstructure noise} \label{sec:microstructure}
\begin{figure}[H]
\begin{tabular}{cc}
  \includegraphics[width=0.45\textwidth]{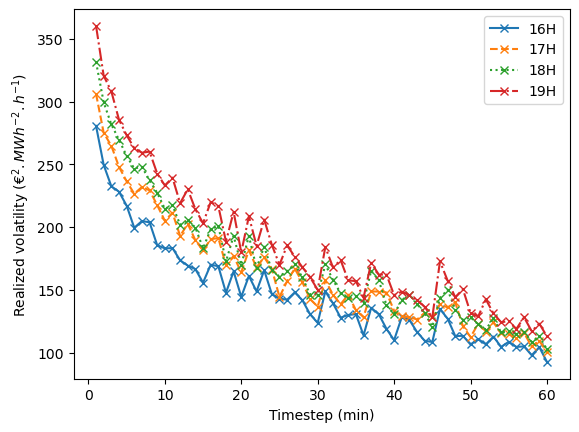}   &  \includegraphics[width=0.45\textwidth]{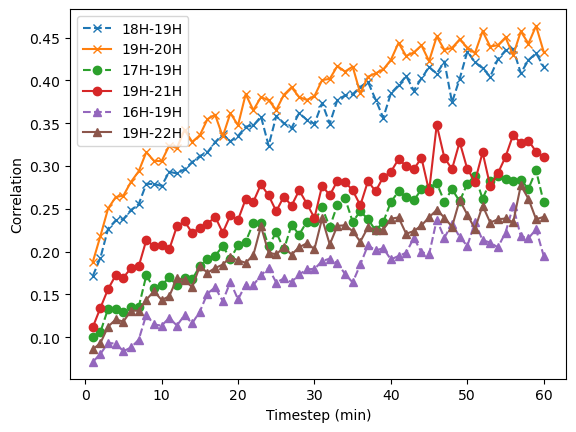} \\
\end{tabular}
    \centering
    \caption{Signature plot \eqref{eq:sig_plot} over $\left[0,T\right]$ normalised by $T^{-1}$ with $T$ being the maturity minus one hour (left) and Epps effect \eqref{eq:correl_estimator} (right) for German transaction prices in 2022 for some delivery periods.}
    \label{fig:sig_plot_epps}
\end{figure}

Here we address the issue of microstructure noise, already identified by Deschatre and Gruet~\cite{deschatre2023b}. Unlike Deschatre and Gruet~\cite{deschatre2023b}, the modelling of this noise is beyond the scope of this paper, but we have to take it into account in the estimation procedure to avoid overestimating the volatility or underestimating the correlation. Two major effects are identified by Deschatre and Gruet~\cite{deschatre2023b}, which are common in the modelling of high-frequency financial data \cite{bacry2013a}:
\begin{itemize}
    \item[(i)] the signature plot of the price over a period $\left[0,T\right]$, $T>0$, shown in Figure~\ref{fig:sig_plot_epps}, is the function 
    \begin{equation} \label{eq:sig_plot}
        \Delta \to \hat{C}(\Delta, T) = \sum_{i=1}^{\lfloor T / \Delta\rfloor} \left(f_{i\Delta} - f_{(i-1)\Delta}\right)^2
    \end{equation}
    for the price $(f_t)_t$, which is the empirical quadratic variation of the price process as a function of the sampling time step. In Figure~\ref{fig:sig_plot_epps}, the curves correspond to the average of \eqref{eq:sig_plot} over the different trading sessions, normalised by $T^{-1}$, $T$ being the maturity minus one hour. For a semi-martingale, this curve should be constant and estimate the integrated volatility. However, we observe an instability when the sampling time step becomes low (frequency becomes high), caused by a mean reverting behaviour of the prices at a high frequency scale. The signature plot then stabilises at lower frequencies, in our case around 30-40 minutes. 
    \item[(ii)] Epps effect is the equivalence of the signature plot in a multidimensional setting. For two prices $(f_{l,t})_t$ and $(f_{m,t})_t$ with maturities $T_l$ and $T_m$, we consider the quadratic covariation estimator over the period $\left[0,T\right]$ 
    \begin{equation} \label{eq:epps}
        \Delta \to \hat{C}_{lm}(\Delta, T) = \sum_{i=1}^{\lfloor T / \Delta\rfloor} \left(f_{l,i\Delta} - f_{l,(i-1)\Delta}\right)\left(f_{m,i\Delta} - f_{m,(i-1)\Delta}\right)
    \end{equation}
    as a function of the sampling time step. When $\Delta$ is small, because prices almost-surely never jump at exactly the same time, the covariance goes to zero. In Figure~\ref{fig:sig_plot_epps}, we plot the correlation estimator 
    \begin{equation} \label{eq:correl_estimator}
    \frac{\sum_{d=1}^D \left(\hat{C}^d_{lm}(\Delta,T_{l,m,e}) - \hat{C}^d_{lm}(\Delta, T_{l,m,b})\right)}{\sqrt{\sum_{d=1}^D \left(\hat{C}^d_{ll}(\Delta,T_{l,m,e}) - \hat{C}^d_{ll}(\Delta, T_{l,m,b})\right)\sum_{d=1}^D \left(\hat{C}^d_{mm}(\Delta,T_{l,m,e}) - \hat{C}^d_{mm}(\Delta,T_{l,m,b})\right)}},
    \end{equation}
 for different maturity pairs $T_l$, $T_m$, where $\hat{C}^d_{lm}$ is the quantity \eqref{eq:epps} estimated for the trading session with delivery date $d \in \{1,\ldots,D\}$ and $\left[T_{l,m,b}, T_{l,m,e}\right]$ is the estimation period, i.e. between 3 pm on the day before delivery and one hour before the minimum between the two maturities. As the sampling time step increases, it increases before stabilising, here around 30-40 minutes. 
\end{itemize} 
The use of Hawkes processes~\cite{bacry2013b} allows these two behaviours to be modelled, but as stated, this is beyond the scope of this paper and could be considered as part of future research. 

\subsection{Correlation} \label{sec:emp_correl}
\begin{figure}[H]
    \centering
    \includegraphics[width=0.9\textwidth]{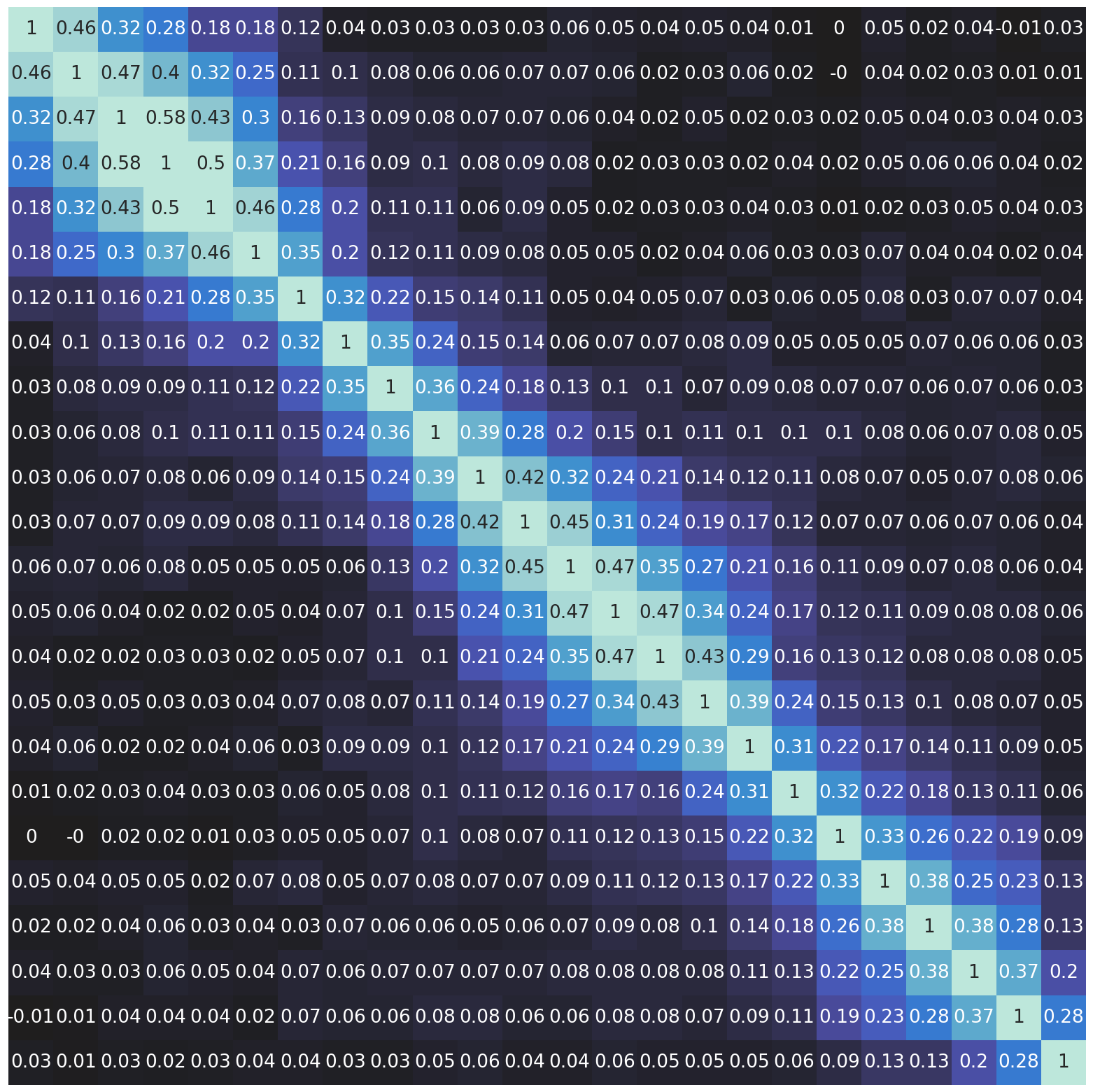} 
    \caption{Correlation matrix estimated using \eqref{eq:correl_estimator} with a sampling time step of 30 minutes for German transaction prices in 2022.}
    \label{fig:correlation_matrix}
\end{figure}

\begin{figure}[H]
    \centering
    \begin{tabular}{cc}
       \includegraphics[width=0.45\textwidth]{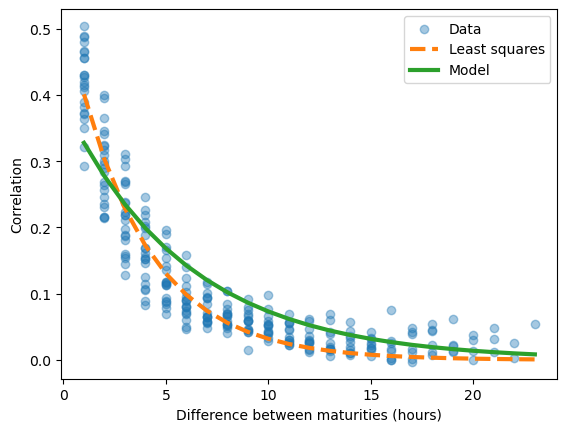}  &  \includegraphics[width=0.45\textwidth]{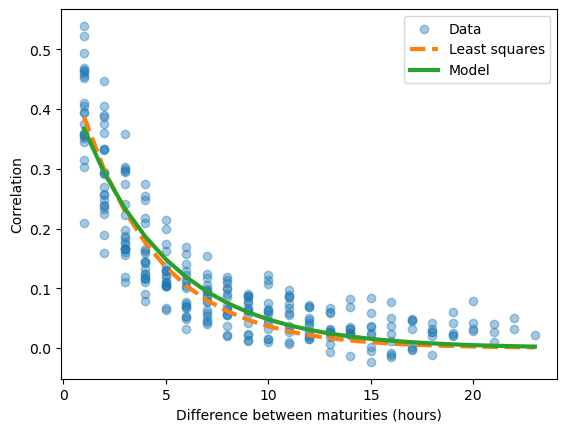}\\
       2019 & 2020 \\
      \includegraphics[width=0.45\textwidth]{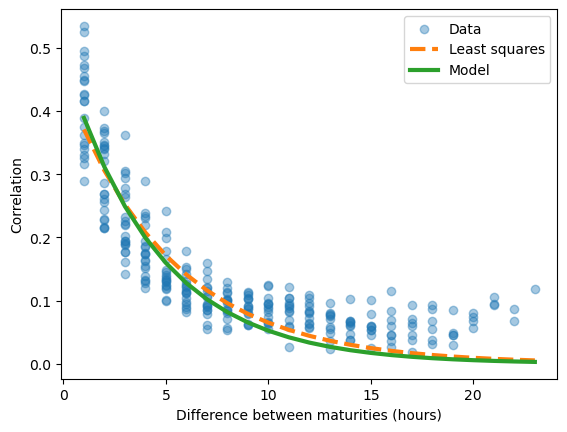}   & \includegraphics[width=0.45\textwidth]{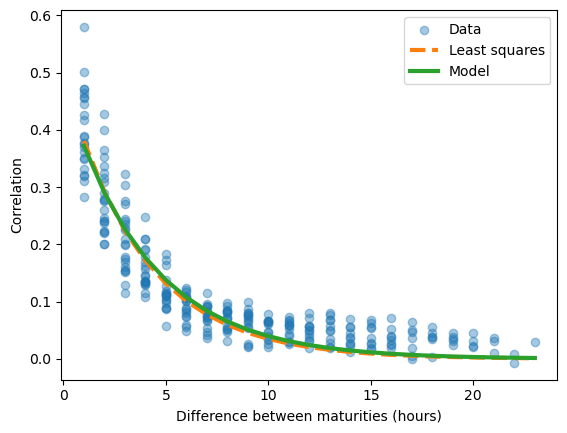}
      \\
      2021 & 2022
    \end{tabular}
    \caption{Correlations as a function of distance between maturities estimated using \eqref{eq:correl_estimator} with a sampling time step of 30 minutes for German transaction prices in 2019, 2020, 2021 and 2022.}
    \label{fig:reg_correl}
\end{figure}

The correlation between different maturities is particularly important for the risk management of assets that are optimised over several delivery hours, such as storage, and is the main stylised fact that we focus on in this paper. The correlation matrix is shown in Figure~\ref{fig:correlation_matrix} for German prices for the year 2022 and the correlation parameters are estimated pairwise using the estimator \eqref{eq:correl_estimator} with a sampling time step of 30 minutes. The correlation between two maturities seems to decrease with the distance between the maturities. This is confirmed by Figure~\ref{fig:reg_correl}, which plots the correlations against the distance between maturities for the years 2019 to 2022. In each year there is an exponential decay with time to maturity: the function $x \to a \exp(\frac{-\kappa}{2}x)$ is fitted to the correlation curve using least squares minimisation. This correlation structure is of particular interest and allows us to propose a sparse model with few parameters. The decrease in correlation as the distance between maturities increases has been identified in future crude oil markets by Schneider and Tavin \cite{schneider2018}; they call this effect the Samuelson correlation effect.

The decrease in correlation and the increase in intensity as the time to maturity increases can be explained by a pure supply-demand equilibrium view. The price will move in the event of a fortuitous event, assuming that traders only buy or sell to balance their position relative to the spot market. For example, if there is a cold snap that was not anticipated at the time of the spot market, consumption will increase and suppliers will have to buy on the market, causing the price to rise. Conversely, if there is more wind than expected, wind producers will have to sell, causing the price to fall. For a given maturity, the closer it is, the greater the probability that the trader will have to buy or sell, or equivalently, the greater the intensity of the price change, as the trader will have to react quickly. If the maturity is further away, the trader can wait for a new forecast. In addition, the uncertainties usually affect several delivery dates at the same time. As in the univariate case, the trader will prefer to buy or sell products with the closest maturities and wait for new forecasts for the most distant maturities. According to this interpretation, the closer the maturities are, the greater the probability that their prices will move simultaneously, and this probability increases as the maturities get closer.

\section{Electricity intraday price modelling}
\label{sec:model}

This section is the core of the paper and presents the price model, its properties and its interpretation. As mentioned above, to the best of our knowledge, this model is the first multidimensional intraday electricity price pure simulation model. We use a point process approach which is natural as we have high frequency asynchronous data, made up of event times associated with values which are prices.  In an equivalent and standard way, we associate upward or downward returns with event times. One of the simplest possible models is to use a non-homogeneous compound Poisson process for each maturity. Non-homogeneity is used to model the increase in intensity and volatility as a function of time, and corresponds to the model of Deschatre and Gruet~\cite{deschatre2023b} without the self-exciting part. However, taking these processes independent of each other does not allow the correlation structure of prices to be modelled. In the following, we create dependence between these inhomogeneous compound Poisson processes using Poisson measures (see Definition~\ref{def:poisson_measure} and \cite[Section 1.1]{bremaud2020} for a general introduction on random and Poisson measures) on the 3-dimensional space $\R_+ \times \R_+ \times K$, with $K$ is the space of jump sizes, i.e. by adding a dimension with respect to a compound Poisson process which can be created from a Poisson measure on the 2-dimensional space $\R_+ \times K$.

\begin{definition}[Poisson random measure, from Definition 2.18 in \cite{tankov03}] \label{def:poisson_measure}
Let $\left(\Omega, \mathcal{F}, \mathbb{P}\right)$ be a probability space, $E \subset \R^d$ and $\mu$ a given (positive) Radon measure on $\left(E,\mathcal{E}\right)$. A Poisson random measure on $E$ with intensity measure $\mu$ is an integer valued random measure:
\[\pi:\Omega \times \mathcal{E} \to \mathbb{N}\]
\[(\omega, A) \mapsto \pi(\omega, A),\]
such that 
\begin{itemize}
    \item[1.] For (almost all) $\omega \in \Omega$, $M(\omega,\cdot)$ is an integer valued Radon measure on $E$ ; for any bounded measurable $A \subset E$, $M(A) < \infty$ is an integer valued random variable.
    \item[2.] For each measurable set $A \subset E$, $M(\cdot, A) = M(A)$ is a Poisson random variable with parameter $\mu(A)$:
    \[\forall k \in \mathbb{N},\, \mathbb{P}\left(M(A)=k\right) = e^{-\mu(A)}\frac{(\mu(A))^k}{k!}.\]
    \item[3.] For disjoint measurable sets $A_1$, $\ldots$, $A_n \in \mathcal{E}$, the variables $M(A_1)$, $\ldots$, $M(A_n)$ are independent. 
\end{itemize}
\end{definition}

\subsection{Model}
Consider $M$ maturities, $0 < T_1<T_2<\cdots<T_M = T$, with $M \in \mathbb{N} \setminus \{0\}$. On a probability space $\left(\Omega, \mathcal{F}, \mathbb{P}\right)$, we consider $2\left(M+1\right)$ independent Poisson measures $\pi_1^+(dt,dx,dy)$, $\ldots$, $\pi_M^+(dt,dx,dy)$, $\pi_1^-(dt,dx,dy)$, $\ldots$, $\pi_M^-(dt,dx,dy)$, $\pi^+(dt,dx,dy)$,  $\pi^-(dt,dx,dy)$ on  $(E, \mathcal{E}) = \left(\R_+ \times \R_+ \times K,\, \mathcal{B}(\R_+) \otimes \mathcal{B}(\R_+) \otimes \mathcal{K}\right)$ with the same intensity measure $ds \otimes dx \otimes \nu(dy)$, where $\nu$ is a probability measure on $\left(K, \mathcal{K}\right)$. $K$ is the space of the price return sizes in absolute value and $\nu$ corresponds to their probability distribution ; hence we assume $K \subset \R_+$ and $\nu(\{0\})=0$ (absolute value of return sizes almost-surely non-negative). As we are interested in volatility and correlations modelling, we also assume that $\int_K y^2 \nu(dy) < \infty$. Note that these Poisson measures correspond to Poisson measures with i.i.d. marks, see \cite[Section 3.3]{bremaud2020}. We endow $\Omega$ with the filtration $(\mathcal{F}_t)_{t \geq 0}$ defined at time $t$ by
{
\scriptsize
\[\bigvee_{h\in\{+,-\}}\left(\sigma\left(\pi^{h}\left( \left(0,s\right] \times B\right):\,0\leq s\leq t,\,B \in \mathcal{B}(\R_+) \otimes \mathcal{K}\right) \bigvee \left( \bigvee_{m=1}^M \sigma\left(\pi^{h}_m\left( \left(0,s\right] \times B\right):\,0\leq s\leq t,\,B \in \mathcal{B}(\R_+) \otimes \mathcal{K}\right)\right)\right).\]
}

\medskip
The $2\left(M+1\right)$ Poisson measures are used to model the $2M$ processes $(f^+_{m,t})_t$ and $(f^-_{m,t})_t$ for $m=1,\ldots,M$ which are respectively the sum of the absolute values of the upward and downward price returns for the price with maturity $T_m$. The model is the following: 

\begin{equation} \label{eq:model_pm}
\begin{split}
f_{m,t}^{h} =& \int_0^t \int_{\R_+} \int_K y {\bf 1}_{x \leq \mu e^{-\kappa(T_m-s)} {\bf 1}_{s \leq T_m}} \pi^{h}_m(ds,dx,dy) \\
&+ \int_0^t \int_{\R_+} \int_K y {\bf 1}_{x \leq \mu_c e^{-\kappa(T_m-s)}{\bf 1}_{s \leq T_m}} \pi^{h}(ds,dx,dy)\,
\end{split}
\end{equation}

for $h=+,-$, $m=1,\ldots,M$ and $0 \leq t \leq T$ with $\mu$, $\mu_c$ and $\kappa \geq 0$. With $f_{m,0}$ the initial price for maturity $T_m$, the price is then given by
\begin{equation}
\label{eq:model}    
f_{m,t} = f_{m,0} + f_{m,t}^{+} - f_{m,t}^{-}, \,0 \leq t \leq T.
\end{equation}
Only three parameters, whose interpretation is given below, can be used to model all the prices, in addition to the law of jumps.

\begin{remark}
    The first term of the sum in \eqref{eq:model_pm} is just an inhomogeneous Poisson process with i.i.d. marks and could simply be defined by a Poisson random measure on $\mathbb{R}_+ \times K$. We keep three dimensional measures to ensure consistency of notation with the second part of the sum.
\end{remark}

\subsubsection*{Marginal properties} First, looking at each process marginally, that is for a given $m \in \{1,\ldots,M\}$, the processes $(f_{m,t}^{+})_t$ and $(f_{m,t}^{-})_t$ are inhomogeneous $\left(\mathcal{F}_t\right)_t$-compound Poisson processes with same intensity at time $0 \leq t \leq T$ which is $\left(\mu + \mu_c\right)e^{-\kappa(T_m-t)} {\bf 1}_{t \leq T_m}$ and jump sizes with law $\nu(dy)$. This result is a direct application of \cite[Theorem 5.7.3]{bremaud2020}. From a martingale point of view, the point processes 
\begin{equation} \label{eq:modelN}
\begin{split}
N_{m,t}^{h} =& \int_0^t \int_{\R_+} \int_K {\bf 1}_{x \leq \mu e^{-\kappa(T_m-s)}{\bf 1}_{s \leq T_m}} \pi^{h}_m(ds,dx,dy) \\
&+ \int_0^t \int_{\R_+} \int_K {\bf 1}_{x \leq \mu_c e^{-\kappa(T_m-s)}{\bf 1}_{s \leq T_m}} \pi^{h}(ds,dx,dy)
\end{split}
\end{equation}
for $h=+,-$ have $(\mathcal{F}_t)_t$-compensator
\[
\begin{split}
&\int_0^t \int_{\R_+} \int_K {\bf 1}_{x \leq \mu e^{-\kappa(T_m-s)}{\bf 1}_{s \leq T_m}} ds dx \nu(dy) + \int_0^t \int_{\R_+} \int_K {\bf 1}_{x \leq \mu_c e^{-\kappa(T_m-s)}{\bf 1}_{s \leq T_m}} ds dx \nu(dy) \\
&= \int_0^t \left(\mu + \mu_c\right) e^{-\kappa(T_m-s)}{\bf 1}_{s \leq T_m}ds
\end{split}
\]
which defines a inhomogeneous Poisson process with intensity $\left(\mu + \mu_c\right) e^{-\kappa(T_m-s)}{\bf 1}_{s \leq T_m}$. The intensity that measures the number of price movements  which is the intensity of $N_{m,t}^{+} + N_{m,t}^{-}$, is then given by $2\left(\mu + \mu_c\right) e^{-\kappa(T_m-s)}{\bf 1}_{s \leq T_m}$ : it increases with time to maturity which is consistent with the empirical results in \cite{deschatre2023b} and Figure~\ref{fig:intensity}. The parameter $\kappa$ represents the rate at which the intensity increases. The intensity becomes 0 after the maturity to represent the end of the trading session for that maturity. Also, the price process $(f_m)_{m \in \{1,\ldots,M\}}$ is a squared integrable martingale, which is a direct consequence of Proposition~\ref{prop:martingale} with proof given in Section~\ref{sec:proofofmartingale}.

\begin{proposition}\label{prop:martingale} In the model~\eqref{eq:model_pm}-\eqref{eq:model}, the two multivariate processes 
\[
\left(f_{m,t}^{h} - \int_{\R} y \nu(dy) \int_0^t \left(\mu + \mu_c\right)e^{-\kappa(T_m-s)}ds\right)_{m=1,\ldots,M},\, t \leq T,\,\,h=+,-,
\]
are squared $(\mathcal{F}_t)_t$-integrable martingales. 
\end{proposition}

\subsubsection*{Dependence} The shared Poisson measures $\pi^{+}$ and $\pi^-$ model the dependence through the processes 
\[f_{m,t}^{c,h} = \int_0^t \int_{\R_+} \int_K y {\bf 1}_{x \leq \mu_c e^{-\kappa(T_m-s)}{\bf 1}_{s \leq T_m}} \pi^{h}(ds,dx,dy),\, 0 \leq t \leq T,\,m=1,\ldots,M,\,h=+,-\]
that are integrals against these same measures. These processes represent the impact of a common shock on the market, such as a power plant shutdown or a change in weather conditions, which can affect the different maturities. The measure $\pi^+$ models shocks that cause price increases (e.g. power plant shutdown) and $\pi^-$ price decreases (e.g. increase in temperature, increase in wind production). The modelling framework is closely related to the Common Poisson Shock Models \cite{powojowski2002, lindskog2003}, which allow to create dependence between Poisson processes. Indeed, considering our model with time-homogeneous intensities (i.e. $\kappa = 0$) leads to 
\[N_{m,t}^{h} = \tilde{N}_{m,t}^{h} + N_t^{c,h},\,0\leq t \leq T, \,m=1,\ldots,M,\,h=+,-,\]
where $\tilde{N}_m^{h}$, $m=1,\ldots,M$, $h=+,-$ are independent Poisson processes with same intensity $\mu$ and $N^{c,h}$ is a Poisson process with intensity $\mu_c$ which is the common shock. The dependence is purely exogenous: joint price movements are induced by exogenous shocks. This approach is quite different from the multivariate Hawkes approach used for prices modelling, where the dependence between prices is purely endogenous: the price of one asset moves because the price of another asset has moved, see \cite{bacry2013a}.

\smallskip
A shock affects the price $m$ at time $t$ with an intensity $\mu_c e^{-\kappa(T_m-t)}{\bf 1}_{t \leq T_m}$ that increases with the time to maturity at the rate $\kappa$: if the trader has information about an event, he is likely to trade at the closest maturities and may not trade at maturities further away as the event can be resolved. The shock does not affect only one maturity, but several at the same time. This interpretation is consistent with the one given in section~\ref{sec:emp_correl}. Proposition~\ref{prop:proba_common_jumps} gives the probability for a set of maturities to have at least one common positive or negative jump between two dates, while another set of maturities does not. For two maturities $T_i < T_j$, if a shock induces a jump for maturity $T_j$, it necessarily induces a jump for maturity $T_i$, while the absence of a jump from a shock at maturity $T_i$ induces the absence of a jump at maturity $T_j$. A shock then starts by affecting the closest maturity and spreads to the maturities that follow. Considering the special case $\mathcal{M}_1 = \{1,\ldots,p\}$ and $\mathcal{M}_2 = \{p+1,\ldots,M\}$ with $p \in \{1,\ldots,M\}$ in Proposition~\ref{prop:proba_common_jumps}, the probability of having a jump only for the first $p$ maturities is of the order of 
\[
\mu_c e^{-\kappa(T_p-u)}\left(1 - e^{-\kappa(T_{p+1}-T_p)}\right)du
\]
on an interval $\left[u, u+du\right]$ with $du \downarrow 0$. Assuming that $T_{p+1} - T_p$ does not depend on $p$, this probability decreases with the number of maturities $p$. The shock causes traders to buy or sell simultaneously at multiple maturities. Traders can trade at $m$ successive maturities, starting with the nearest one, with probability decreasing with $m$. The proof of Proposition~\ref{prop:proba_common_jumps} is given in Section~\ref{sec:proofofprobacommonjumps}. Note that a common shock induces a jump of the same size for each maturity, which is a strong assumption.

\begin{proposition}\label{prop:proba_common_jumps}
With $\mathcal{M}_1$, $\mathcal{M}_2 \subset \{T_1,\ldots,T_M\}$ and $0 \leq u \leq t \leq \min \mathcal{M}_1 \cup \mathcal{M}_2$, we have for $h=+,-$,
\[
\begin{split}
\mathbb{P}&\left(\int_u^t \int_{\R_+} \int_K {\bf 1}_{x \leq \mu_c e^{-\kappa(T_m-s)}} \pi^{h}(ds,dx,dy) \geq 1 \text{ for every }m \in \mathcal{M}_1 \text{ and } \right.\\
&\left. \int_u^t \int_{\R_+} \int_K {\bf 1}_{x \leq \mu_c e^{-\kappa(T_m-s)}} \pi^{h}(ds,dx,dy) = 0\text{ for every }m \in \mathcal{M}_2 \right)
\end{split}
\]
is equal to
\[\left(1-\exp\left(-\int_{u}^t \mu_c \left(e^{-\kappa(T_{\max \mathcal{M}_1}-s)} - e^{-\kappa(T_{\min \mathcal{M}_2}-s)}\right)ds\right)\right) \exp\left(-\int_{u}^t \mu_c e^{-\kappa(T_{\min \mathcal{M}_2}-s)}ds\right)
\]
if $\max \mathcal{M}_1 < \min \mathcal{M}_2$ and is null otherwise.
\end{proposition}
\begin{remark} By considering the set of maturities as a continuous rather than a discrete set, with for example $T \in \left[T_1, T_2\right]$, $T_1$, $T_2 > 0$ (which is common in the literature devoted to modelling forward prices in electricity markets \cite{deschatre2021}), and with $N^{c,h}_t(T) = \int_0^t \int_{\R_+} \int_K {\bf 1}_{x \leq e^{-\kappa(T-s)}{\bf 1}_{s \leq T}} \pi^{h}(ds,dx,dy)$ for $h=+,-$ and $t \leq T_2$, the common shock component (omitting the marks and normalising by $\mu_c$), the `spot' process $N^{c,h}_T(T)$, (taking the limit $t \to T$ is also common to obtain the spot price from the forward price) is a Trawl process with Trawl $\{(x,s): \, s \leq 0,\, 0 \leq x \leq \exp(\kappa s)\}$. It is interesting to note that these processes, introduced by Barndorff-Nielsen in \cite{barndorff2011}, are used to model spot electricity prices in \cite{veraart2023}.
\end{remark}

\subsection{Covariance matrix}

\medskip
In this section, we focus on the moments of order 2 that correspond to the quadratic covariation of the process. Its empirical counterpart is given by 
\begin{equation} \label{eq:empirical_sig_plot}
\hat{C}_{kl}\left(\Delta, T\right) = \sum_{i=1}^{\lfloor \frac{T}{\Delta} \rfloor} \left(f_{k,i\Delta} - f_{k,(i-1)\Delta}\right) \left(f_{l,i\Delta}- f_{l,(i-1)\Delta}\right), \;k,l=1,\ldots,M
\end{equation}
for $T > \Delta > 0$, $\Delta$ being the sampling time step. The expectation of \eqref{eq:empirical_sig_plot} is given in Proposition~\ref{prop:moment_2}.

\begin{proposition} \label{prop:moment_2} We have
\[\E\left(\hat{C}_{kl}(\Delta,T)\right) = 2 \int_{K} y^2 \nu(dy) \int_{0}^{\Delta \lfloor \frac{T}{\Delta} \rfloor} \left(\mu + \mu_c \delta_{kl}\right) e^{-\kappa \left(\max(T_k,T_l) - s\right)}{\bf 1}_{s \leq \min(T_k,T_l)}ds\]
for $k,l=1,\ldots,M$, $T > \Delta > 0$ and $\hat{C}_{kl}\left(\Delta,T\right)$ given by \eqref{eq:empirical_sig_plot}.
\end{proposition}

The proof follows from the computation of $\E\left( \left(f_{k,i\Delta} - f_{k,(i-1)\Delta}\right) \left(f_{l,i\Delta}- f_{l,(i-1)\Delta}\right) \right)$ which is derived from \cite[Theorem 3.2.1]{bremaud2020} and is equal to 
\[
2\int_{K} y^2 \nu(dy) \int_{(i-1)\Delta}^{i\Delta} \left(\mu_c + \mu \delta_{kl}\right) e^{-\kappa \left(\max(T_k,T_l) - s\right)}{\bf 1}_{s \leq \min(T_k,T_l)}ds.
\]

\medskip
When $T$ is a multiple of $\Delta$, we get
\[ \E\left(\hat{C}_{kl}(\Delta,T)\right) = 2 \int_{K} y^2 f(dy) \int_{0}^{T} \left(\mu_c + \mu \delta_{kl}\right) e^{-\kappa \left(\max(T_k,T_l) - s\right)}{\bf 1}_{s \leq \min(T_k,T_l)}ds.
\]
As this quantity does not depend on $\Delta$, the model is unable to reproduce the signature plot and the Epps effect identified in \cite{deschatre2023b} or in Figure~\ref{fig:sig_plot_epps}, i.e. quadratic covariation depending on the sampling time step. This limitation of the model needs to be taken into account in the estimation procedure. One way to account for these effects is to add self- and cross-excitation in the point processes as in \cite{deschatre2023b} for the univariate case, but this is beyond the scope of this paper.

\smallskip
A proxy for the squared integrated volatility of the price $(f_{m,t})_t$, $m \in \{1,\ldots,M\}$ between 0 and $t$ is then $\mathbb{E}\left(\hat{C}_{mm}(\Delta,t)\right)$ when $\Delta \to 0$ equal to 
 \begin{equation} \label{eq:integrated_volatility} 2\int_K y^2 \nu(dy) \int_0^t \left(\mu + \mu_c\right)e^{-\kappa \left(T_m - s\right)}{\bf 1}_{s \leq T_m}ds.
 \end{equation}
The instantaneous squared volatility at time $t$ can then be considered as the derivative of \eqref{eq:integrated_volatility} equal to 
 \begin{equation} \label{eq:proxy_inst_vol}
 2\int_K y^2 \nu(dy) \left(\mu + \mu_c\right)e^{-\kappa \left(T_m - t\right)}{\bf 1}_{t \leq T_m}
 \end{equation}
which is the instantaneous intensity of price changes times the order 2 moment of the jump sizes. The notions of intensity and volatility are closely related, and the Samuelson effect applies to both, as noted above.

\smallskip
The correlation between $f_k$ and $f_l$ for $k \neq l$ is given by 
\begin{equation} \label{eq:correl_th}
\frac{\E\left(\hat{C}_{kl}(\Delta,t)\right)}{\sqrt{\E\left(\hat{C}_{kk}(\Delta,t)\right)\E\left(\hat{C}_{ll}(\Delta,t)\right)}} = \frac{\mu_c}{\mu + \mu_c}e^{-\frac{\kappa}{2} \left|T_l - T_m\right|},\;t \leq \min(T_k,T_l).
\end{equation} 
The correlation decreases as the maturity distance increases, which is consistent with the empirical results of Section~\ref{sec:empirical_facts}. The rate of decrease in correlation, $\frac{\kappa}{2}$, is half the rate of increase in intensity, i.e. $\kappa$, and equal to the rate of increase in volatility (not squared). The parameter $\kappa$ in our model represents both the Samuelson effect and the Samuelson correlation effect. Moreover, the correlation in our model does not depend on the time $t$ ; this property is in line with the results of Hirsch and Ziel~\cite{hirsch2023}, who does not obtain an increase in forecast quality by implementing a time-dependent dependency structure.

\medskip
The model can represent both the Samuelson effect and the correlation structure with only three parameters (without considering the law of the jumps), which is its main advantage.

\subsection{Simulation method}

\begin{figure}[H]
    \centering
    \includegraphics[width=0.49\textwidth]{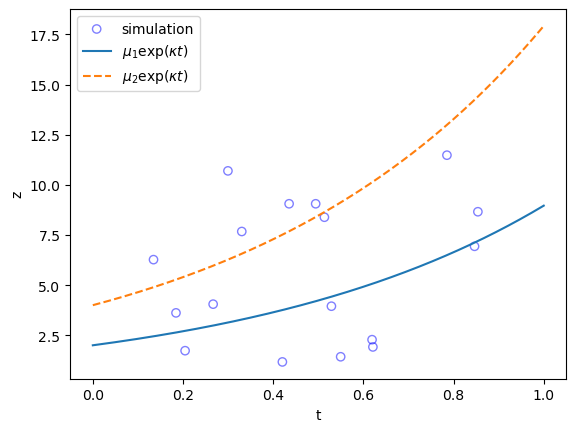}
    \caption{A simulation of a homogeneous Poisson measure with intensity $dt \otimes dx$ on $\left[0,1\right] \times \left[0,\max_{i=1,2} \sup_{s\in\left[0,t\right]} f_i(s)\right]$ with $f_1(t) = \mu_1 \exp(\kappa t)$ and $f_2(t) = \mu_2 \exp(\kappa t)$, $\mu_1 = 2$, $\mu_2 = 4$ and $\kappa = 1.5$. For this realisation, $N^{f_1}_1 = 7$ and $N^{f_2}_1 = 12$, where $N^f$ is defined in \eqref{eq:poisson_measure_f}.}
    \label{fig:poisson_measure}
\end{figure}

It is possible to simulate processes of the form 
\begin{equation} \label{eq:poisson_measure_f}
N^{f_i}_t = \int_0^t \int_{\R_+} {\bf 1}_{z \leq f_i(s)}\pi(ds,dz),\,i=1,\ldots,I
\end{equation}
that are present in the coupling part of the prices using \cite[Theorem 3.1.1]{bremaud2020} in the following way:
\begin{itemize}
    \item[(i)] first simulate the number of jumps $N$ of a homogeneous Poisson measure on \\ $\left[0,t\right] \times \left[0,\max_{i=1,\ldots,I} \sup_{s\in\left[0,t\right]} f_i(s)\right]$, which follows a Poisson law with parameter $t \times \max_{i=1,\ldots,I} \sup_{s\in\left[0,t\right]} f_i(s)$ ;
    \item[(ii)] then simulate the $N$ time jumps $T_1,\ldots,T_N$ with uniform law on $\left[0,t\right]$ and the $N$ coordinates in $x$, $X_1,\ldots,X_N$ with uniform law on $\left[0,\max_{i=1,\ldots,I} \sup_{s\in\left[0,t\right]} f_i(s)\right]$ ;
    \item[(iii)] for each $i=1,\ldots,I$, we have $N^{f_i}$ equal to the number of jumps such that  $X_n \leq f_i(T_n)$, $n=1,\ldots,N$.
\end{itemize}
A simulation of a Poisson measure on $\mathbb{R}_+^2$ with the simulation of $N_t^{f_i}$, $i=1,2$, is displayed in Figure \ref{fig:poisson_measure}.

Proposition~\ref{prop:simulation} gives an alternative simulation method: the coupling part of the prices can be written as a linear combination of compound Poisson processes. The proof is given in Section~\ref{sec:proofofsimulation}. The idea of the proof is to divide the space into disjoint regions, the one below the blue curve and the one between the solid blue curve and the dashed orange curve in Figure~\ref{fig:poisson_measure}, using the property $\mu_c e^{-\kappa(T_i-s)} > \mu_c e^{-\kappa(T_j-s)}$ for $T_i > T_j$.

\begin{proposition} \label{prop:simulation}Assume $\mu_c > 0$ and $\kappa > 0$. For $h=+,-$, we consider a vector of $M$ independent compound Poisson processes, $P^{c,h}$, with intensities $\mu_c e^{-\kappa\left(T_1-s\right)}$ for $P^{c,h}_1$ and $\mu_c e^{-\kappa\left(T_m-s\right)} - \mu_c e^{-\kappa\left(T_{m-1}-s\right)}$ for $P^{c,h}_m$, $m=2,\ldots,M$ and with the jump law $\nu(dy)$. With 
\[f^{c,h}_t = \left(\int_0^t \int_{\R_+} \int_K y {\bf 1}_{x \leq \mu_c e^{-\kappa\left(T_m-s\right)}{\bf 1}_{s \leq T_m}} \pi^{h}\left(ds, dx, dy\right)\right)_{m=1,\ldots,M},\,0\leq t \leq T,\] we have 
\[f^{c,h} \underset{law}{=} \left(\int_0^{t} H_s dP^{c,h}_s\right)_t\]
with $H_s$ the matrix of size $M \times M$ such that $H_{ij,s} = {\bf 1}_{j \leq i}{\bf 1}_{s \leq T_i}$ for $i,j=1,\ldots,M$.
\end{proposition}

Corollary~\ref{cor:simulation} is a direct consequence of Proposition~\ref{prop:simulation} and shows that we can simulate the price process using $4M$ independent Poisson processes.
\begin{corollary}\label{cor:simulation}Assume $\mu > 0$, $\mu_c > 0$, and $\kappa > 0$. We have
\[(f_t)_t \underset{law}{=} \left(\int_0^{t} I_s d\left(P^{+}-P^{-}\right)_s + \int_0^{t} H_s d \left(P^{c,+}-P^{c,-}\right)_s\right)_t\]
where $P^{+}_1,\,P^{-}_1,\ldots,P^{+}_M,\,P^{-}_M,\,P^{c,+}_1,\,P^{c,-}_1,\ldots,P^{c,+}_M,\,P^{c,-}_M$ are $4M$ independent compound Poisson processes with intensities 
\begin{itemize}
    \item[-] $\mu e^{-\kappa(T_m-s)}$ for $P^{+}_m$ and $P^{-}_m$, $m \geq 1$,
    \item[-] $\mu_c e^{-\kappa(T_1-s)}$ for $P^{c,+}_1$ and $P^{c,-}_1$,
    \item[-] $\mu_c e^{-\kappa(T_m-s)} - \mu_c e^{-\kappa(T_{m-1}-s)}$ for $P^{c,+}_m$ and $P^{c,-}_m$, $m \geq 2$,\end{itemize}
and jump law $\nu(dy)$, $H_s$ and $I_s$ are the matrices of size $M \times M$ such that $H_{ij,s} = {\bf 1}_{j \leq i}{\bf 1}_{s \leq T_i}$ and $I_{ij,s} = \delta_{ij}{\bf 1}_{s \leq T_i}$ for $i,j=1,\ldots,M$. 
\end{corollary}

\subsection{Limit theorems}
\label{sec:limit_theorems}
In this section we consider an asymptotic setting on the intensity parameters $\mu$ and $\mu_c$ growing with some $n \in \mathbb{N} \setminus \{0\}$ while keeping $T$ fixed, as in \cite{fromont2011, chen2013, chen2016, deschatre2023}: $\mu = \mu^{(n)} = n\tilde{\mu}$, $\mu_c = \mu_c^{(n)} = n\tilde{\mu}_c$, with $\tilde{\mu},\,\tilde{\mu}_c > 0$ fixed. This allows us to study the convergence properties of the price process as $n \to \infty$. The price process is now denoted by $(f^{(n)}_t)_t$ to indicate the dependence on $n$. This asymptotic setting corresponds to three different statistical frameworks:
\begin{itemize}
\item[(i)] the number of observed jumps is high: indeed, as $n$ increases, the intensity of the different point processes and then the number of jumps increases;
    \item[(ii)] $n$ is the number of i.i.d. observations (for instance the number of observed trading sessions) of the price process $(\tilde{f})_t$ with parameters $\kappa$, $\tilde{\mu}$ and $\tilde{\mu}_c$, $(\tilde{f}^{i})_t$, $i=1,\ldots,n$ and $f^{(n)}$ corresponds to the superposition of the different price processes, that is $(f_t^{(n)})_t = (\sum_{i=1}^n \tilde{f}_t^{i})_t$;
    \item[(iii)] $f^{(n)}_{t} = \tilde{f}_{tn}$, $t \in \left[0, T\right]$ with $(\tilde{f}_t)_t$ a process defined in the same way than $(f_{t})_t$, observed on $\left[0, nT\right]$ but with time dependent intensities $\mu e^{-\kappa(T_m - \frac{t}{n})}$ and $\mu_c e^{-\kappa(T_m - \frac{t}{n})}$ for $m=1,\ldots,M$: the asymptotic framework is the same one that a price process observed on a large time horizon $\left[0, nT\right]$.
\end{itemize}

\begin{remark} Since from Corollary~\ref{cor:simulation}, the price process $(f_t)_t$ is a linear combination of independent compound Poisson processes, the equivalence between (i), (ii), (iii) is immediate and is standard for Poisson processes.
\end{remark}
\

Proposition~\ref{prop:moment1conv} gives convergence results about the moments of order 1 when $n \to \infty$. In particular, taking $g(y)=y$ gives the convergence of $(f_t^{h,(n)})_t$, $h=+,-$. The proof is given in Section~\ref{sec:proofofmoment1conv}.

\begin{proposition} \label{prop:moment1conv} With $g(y)$ a measurable function on $\left(K,\mathcal{K}\right)$ such that $\int_K g^2(y)\nu(dy) < \infty$ and $f^{g,h, (n)}_t$ defined by
\[
\begin{split}
&\left(\int_{0}^t \int_{\R_+} \int_{K}g(y){\bf 1}_{x \leq n\tilde{\mu} e^{-\kappa(T_m-s)}{\bf 1}_{s \leq T_m}} \pi_m^{h}(ds,dx,dy) + \right.\\
&\left.\int_{0}^t \int_{\R_+} \int_{K}g(y){\bf 1}_{x \leq n\tilde{\mu}_c e^{-\kappa(T_m-s)}{\bf 1}_{s \leq T_m}} \pi^{h}(ds,dx,dy)\right)_{m=1,\ldots,M}
\end{split}
\]
for $t \leq T$ and $h=+,-$, we have
\[\underset{t\in\left[0,T\right]}{\sup}\left\Vert n^{-1}f^{g,h,(n)}_t - \left(\int_{K}g(y)\nu(dy) \int_{0}^t \left(\tilde{\mu} + \tilde{\mu}_c\right)e^{-\kappa\left(T_m-s\right)} {\bf 1}_{s \leq T_m}ds\right)_{m=1,\ldots,M}\right\Vert\to 0\]
in $L^2(\mathbb{P})$ as $n \to \infty$.
\end{proposition}

Proposition~\ref{prop:trajectoryconv} gives the limit of the scaled martingale $(\frac{1}{\sqrt{n}}f^{(n)}_t)_t$ as $n \to \infty$. The limit is a process driven by a multivariate Brownian motion with correlation decreasing when distance between maturities increases and with volatility increasing with time to maturity at the rate $\kappa$. The instantaneous volatility in the limit diffusion is the same as in Equation~\eqref{eq:proxy_inst_vol} (to within one normalisation). The proof is given in Section~\ref{sec:proofoftrajectoryconv}.
The two quantities of interest highlighted in Proposition~\ref{prop:trajectoryconv} are 
\begin{equation} 
\label{eq:vol_param}
\sigma = \sqrt{\frac{2\left(\mu + \mu_c\right)}{\kappa}\int_{K}y^2 \nu(dy)} \left(\approx \sqrt{2\left(\mu + \mu_c\right)\int_{K}y^2 \nu(dy) \int_0^{T_m} e^{-\kappa(T_m-s)}ds} \right)
\end{equation}
which is a proxy for the square root of the integrated volatility and 
\begin{equation} \label{eq:correl_param}
\rho = \frac{\mu_c}{\mu + \mu_c}e^{-\frac{\kappa}{2} \Delta T}
\end{equation}
with $\Delta T$ equal to 1 hour, which corresponds to the correlation between two products with consecutive maturities. To obtain the correlation between two products with maturities separated by $r$ hours, simply calculate $\rho e^{-\frac{\kappa}{2}(r-1)\Delta T}$.

\begin{proposition} \label{prop:trajectoryconv} If $\int_K y^4 \nu(dy) < \infty$, the process $\frac{1}{\sqrt{n}}f^{n}_t,\,t \in \left[0, T\right]$ converges in law for the Skorokhod topology to 
\[
\left(\int_0^t \sqrt{\int_K y^2 \nu(dy)}\sqrt{2\left(\tilde{\mu} + \tilde{\mu}_c\right)} e^{-\frac{\kappa}{2}(T_m - s)}{\bf 1}_{s \leq T_m} dW_{m,s}\right)_{m=1,\ldots,M},\,t \in \left[0, T\right]
\]
as $n \to \infty$ where $W = \left(W_1, \ldots,W_M\right)^{\top}$ is a multivariate Brownian motion with correlation matrix $\left(\frac{\tilde{\mu} \delta_{kl} + \tilde{\mu_c}}{\tilde{\mu} + \tilde{\mu}_c}e^{-\frac{\kappa}{2}|T_k-T_l|}\right)_{k,l=1,\ldots,M}$.
\end{proposition}

\section{Estimation and numerical results}
\label{sec:estimation}
Here, we present a moment-based procedure for estimating the parameters from the data, and the numerical results that follow. This procedure is particularly simple and easy to implement.

\subsection{Estimation procedure}

For each maturity $T_m$, $m=1,\ldots,M$, we estimate the model only on a portion of the trading session $\left[T_{b,m}, T_{e,m}\right] \subset \left[0, T_m\right]$ and we have a sample of $D$ days, $D \in \mathbb{N} \setminus \{0\}$. 

\subsubsection*{Jump size law} Note that we observe the empirical distribution of the jump sizes because the price processes are observed continuously. Assuming that the jump size law $\nu(dy)$ is discrete, which is consistent with the fact that the price lives on a discrete grid and is proportional to the tick size equal to 0.01\euro{}/MWh in our data, we estimate the different probabilities of this discrete law with the empirical probabilities, taking into account both negative and positive jumps in absolute value as well as all maturities. Let $\hat{\nu}$ the empirical distribution that estimates $\nu$.

\subsubsection*{Estimation of $\kappa$} The parameter $\kappa$ represents the rate at which the intensity of price changes increases. To estimate this parameter, we use the least squares estimator that minimises the contrast
\begin{equation} \label{eq:least_square_init}
 \tilde{\mathcal{L}}_m= -4\left(\mu + \mu_c\right) \int_{T_{b,m}}^{T_{e,m}} e^{-\kappa (T_m - s)}d\left(N_{m,s}^{+} + N_{m,s}^{-}\right) + 4\left(\mu + \mu_c\right)^2 \int_{T_{b,m}}^{T_{e,m}} e^{-2\kappa (T_m - s)}ds
\end{equation}
with $N_{m,s}^{+}$ and $N_{m,s}^{-}$ defined by \eqref{eq:modelN}. This estimator has been used, for example, in the context of inhomogeneous Poisson process intensity estimation by Reynaud-Bouret \cite{reynaud2003}. This contrast allows one to estimate $\kappa$ but also $\left(\mu + \mu_c\right)$. The intensity is strongly related to the volatility: for two Poisson processes $(N_t^+)_t$ and $(N_t^-)_t$, when the sampling time step is close to 0, the instantaneous volatility of the price $N^+_t - N^-_t$ is $\left(dN_s^{+} - dN_s^-\right)^2 \approx dN_s^{+} + dN_s^-$, which is the intensity. Estimating $\mu + \mu_c$ from the contrast \eqref{eq:least_square_init} leads to an overestimation of the volatility: $\mu + \mu_c$ is fitted to represent the intensity, which is the first point of the signature plot (down to the size of the jumps) in Figure~\ref{fig:sig_plot_epps}. In our model, the signature plot is flat and independent on the sampling time step: the macroscopic volatility we are interested in (the battery is optimised with a time step of one hour) is overestimated. This problem highlights a limitation of our model, which is that it cannot model both intensity (equivalent to microscopic volatility) and volatility (macroscopic). The introduction of self-excitation by Hawkes processes as in~\cite{deschatre2023b} could solve this problem, but requires a more complex modelling framework. To address this issue here, we estimate $\kappa$ by parameterizing $\tilde {\mathcal{L}}_m$ using
\[
\begin{split}
\Lambda_m &= \frac{\E\left(N_{m,T_{e,m}}^{+} + N_{m,T_{e,m}}^{-} + N_{m,T_{b,m}}^{+} + N_{m,T_{b,m}}^{-}\right)}{T_{e,m} - T_{b,m}} \\
&= 2\left(T_{e,m} - T_{b,m}\right)^{-1}\left(\mu + \mu_c\right) \int_{T_{b,m}}^{T_{e,m}} e^{-\kappa\left(T_m-s\right)}ds
\end{split}
\] 
the average number of jumps per unit of time. This leads to :
\begin{equation}\label{eq:least_square_new}
\begin{split} 
\tilde {\mathcal{L}}_m = &-2\left(T_{e,m} - T_{b,m}\right)\Lambda_m \left(\int_{T_{b,m}}^{T_{e,m}} e^{-\kappa\left(T_m-s\right)}ds\right)^{-1}\int_{T_{b,m}}^{T_{e,m}} e^{-\kappa (T_m - s)}d\left(N_{m,s}^{+} + N_{m,s}^{-}\right) \\
&+ \left(T_{e,m} - T_{b,m}\right)^2\Lambda_m^2 \left(\int_{T_{b,m}}^{T_{e,m}} e^{-\kappa\left(T_m-s\right)}ds\right)^{-2}  \int_{T_{b,m}}^{T_{e,m}} e^{-2\kappa (T_m - s)}ds
\end{split}
\end{equation}
and depends only on $\kappa$. The empirical counterpart of \eqref{eq:least_square_new} is given by 
\begin{equation}\label{eq:empirical_ls_intensity}
\begin{split}
\hat{\mathcal{L}}_m(\kappa) =&  -2\left(T_{e,m} - T_{b,m}\right)\hat{\Lambda}_m \left(\int_{T_{b,m}}^{T_{e,m}} e^{-\kappa\left(T_m-s\right)}ds\right)^{-1}D^{-1}\sum_{d=1}^D\int_{T_{b,m}}^{T_{e,m}} e^{-\kappa (T_m - s)}d\left(N_{m,s}^{d,+} + N_{m,s}^{d,-}\right) \\
&+ \left(T_{e,m} - T_{b,m}\right)^2\left(\hat{\Lambda}_m\right)^2 \left(\int_{T_{b,m}}^{T_{e,m}} e^{-\kappa\left(T_m-s\right)}ds\right)^{-2} \int_{T_{b,m}}^{T_{e,m}} e^{-2\kappa (T_m - s)}ds
\end{split}
\end{equation}
where $N_m^{d,+}$ and $N_m^{d,-}$ are the observations of $N_m^{+}$ and $N_m^{-}$ for trading session $d$, and $\hat{\Lambda}_m$ an estimate of $\Lambda_m$ equal to the total number of jumps across all sessions between $T_{b,m}$ and $T_{e,m}$ divided by $D \times (T_{e,m} - T_{b,m})$. Since $\kappa$ is the same for each maturity, we consider the estimator $\hat{\kappa}$ that minimises 
\begin{equation} \label{eq:sum_ls}
\sum_{m=1}^M \hat{\mathcal{L}}_m(\kappa).
\end{equation}

\subsubsection*{Estimation of $\mu + \mu_c$} $\mu + \mu_c$ is a proxy for the macroscopic variance and can be estimated using the estimator $D^{-1} \sum_{d=1}^D \hat{C}^d_{mm}(\Delta, T_{m,e}) - \hat{C}^d_{mm}(\Delta,T_{m,b})$, where $\hat{C}^d_{mm}(\Delta, T_{m,e})$ is an estimate of the empirical variance with a time step $\Delta$ for the sample $d$ given by \eqref{eq:empirical_sig_plot}. Considering $\Delta$ large enough to take into consideration microstructure noise, $D^{-1} \sum_{d=1}^D \hat{C}^d_{mm}(\Delta, T_{m,e}) - \hat{C}^d_{mm}(\Delta, T_{m,b})$ estimates
\[
2 \int_{K} y^2 \nu(dy) \int_{\lfloor T_{m,b}/\Delta \rfloor \Delta}^{\lfloor T_{m,e}/\Delta \rfloor \Delta} \left(\mu_c + \mu\right) e^{-\kappa \left(T_m - s\right)}ds,
\]
see Proposition~\ref{prop:moment_2}. An estimator for $\mu + \mu_c$ is then given by $\hat{\mu}_S$ that minimizes
\[
\sum_{m=1}^M \left(D^{-1} \sum_{d=1}^D \left(\hat{C}^d_{mm}(\Delta, T_{m,e}) - \hat{C}^d_{mm}(\Delta,T_{m,b} )\right) - 2 \mu_S \int_K y^2 \hat{\nu}(dy)\int_{\lfloor T_{m,b}/\Delta \rfloor \Delta}^{\lfloor T_{m,e}/\Delta \rfloor \Delta}  e^{-\hat{\kappa} \left(T_m - s\right)}ds\right)^2
\]
and is equal to
\[
\hat{\mu}_S = \frac{\sum_{m=1}^M D^{-1} \sum_{d=1}^D \left(\hat{C}^d_{mm}(\Delta, T_{m,e}) - \hat{C}^d_{mm}(\Delta, T_{m,b})\right) \int_{\lfloor T_{m,b}/\Delta \rfloor \Delta}^{\lfloor T_{m,e}/\Delta \rfloor \Delta} e^{-\hat{\kappa} \left(T_m - s\right)}ds  }{2\int_K y^2 \hat{\nu}(dy)\sum_{m=1}^M \left(\int_{\lfloor T_{m,b}/\Delta \rfloor \Delta}^{\lfloor T_{m,e}/\Delta \rfloor \Delta}  e^{-\hat{\kappa} \left(T_m - s\right)}ds\right)^2}.
\]

\subsubsection*{Estimation of $\frac{\mu_c}{\mu + \mu_c}$} For two maturities $T_l$ and $T_m$, we estimate every day the quadratic covariation matrix \eqref{eq:empirical_sig_plot} on the interval $\left[T_{b,l,m},T_{e,l,m}\right] = \left[T_{b,l},T_{e,l}\right] \cap \left[T_{b,m},T_{e,m}\right]$ (i.e. $T_{b,l,m} = \max(T_{b,l},T_{b,m})$ and $T_{e,l,m} = \min(T_{e,l}, T_{e,m})$ if $\left[T_{b,l,m},T_{e,l,m}\right] \neq \emptyset$) with a sampling time step $\Delta$ large enough (to account for microstructure noise). We only consider maturities such that $T_{e,l,m} - T_{b,l,m} \geq \delta$. To estimate  
\[
\frac{\mu_c}{\mu + \mu_c} e^{-\frac{\kappa}{2}|T_l - T_m|},
\]
we use the empirical correlation
\begin{equation} \label{eq:criteria_correl}
\hat{\rho}_{lm} = \frac{\sum_{d=1}^D \left(\hat{C}^d_{lm}(\Delta, T_{l,m,e}) - \hat{C}^d_{lm}(\Delta, T_{l,m,b})\right)}{\sqrt{\sum_{d=1}^D \left(\hat{C}^d_{ll}(\Delta, T_{l,m,e}) - \hat{C}^d_{ll}(\Delta, T_{l,m,b})\right)\sum_{d=1}^D \left(\hat{C}^d_{mm}(\Delta, T_{l,m,e}) - \hat{C}^d_{mm}(\Delta, T_{l,m,b})\right)}},
\end{equation}
see \eqref{eq:correl_th}. An estimator of $\mu_R = \frac{\mu_c}{\mu + \mu_c}$, $\hat{\mu}_R$, is then given by the minimisation of
\[\sum_{l,m=1,\ldots,M | T_{e,l,m} - T_{b,l,m} \geq \delta} \left(\hat{\rho}_{lm} - \mu_R e^{-\frac{\hat{\kappa}}{2}|T_l - T_m|}  \right)^2,
\]
under the constraint $\mu_R \leq 1$, hence
\[
\hat{\mu}_R = \min\left(\frac{\sum_{l,m=1,\ldots,M | T_{e,l,m} - T_{b,l,m} \geq \delta} \hat{\rho}_{lm} e^{-\frac{\hat{\kappa}}{2}|T_l - T_m|}}{\sum_{l,m=1,\ldots,M | T_{e,l,m} - T_{b,l,m} \geq \delta}e^{-\hat{\kappa}|T_l - T_m|}},1\right).
\]

\medskip
From $\hat{\mu}_S$ and $\hat{\mu}_R$, we estimate $\mu_c$ by $\hat{\mu}_c = \hat{\mu}_S \hat{\mu}_R$ and $\mu$ by $\hat{\mu} = \hat{\mu}_S - \hat{\mu}_c$.

\subsection{Numerical results}
\label{sec:numerical_results}
\begin{figure}[H]
    \centering
    \begin{tabular}{cc}
    \includegraphics[width=0.45\textwidth]{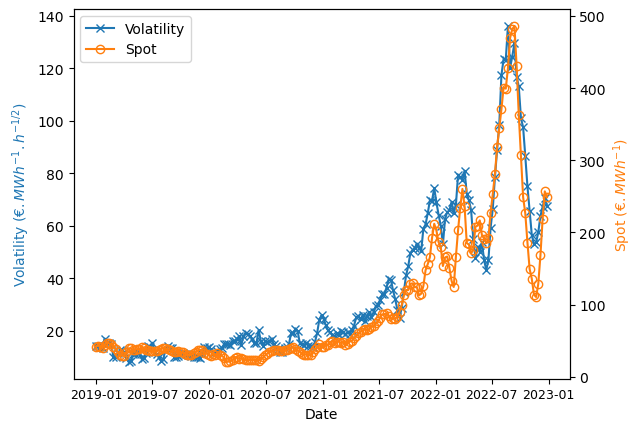} & \includegraphics[width=0.45\textwidth]{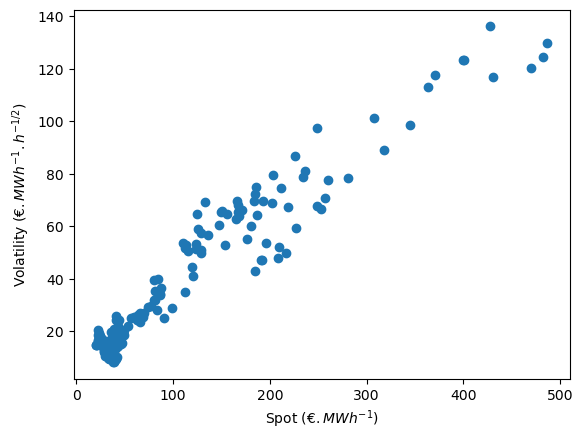}
    \end{tabular}
    \caption{Volatility proxy parameter \eqref{eq:vol_param} estimated each week from the last 28 trading sessions of the German intraday market against the average German spot price during the estimation period.}
    \label{fig:vol_spot_de}
\end{figure}

\begin{figure}[H]
   \centering
    \begin{tabular}{cc}
    \includegraphics[width=0.45\textwidth]{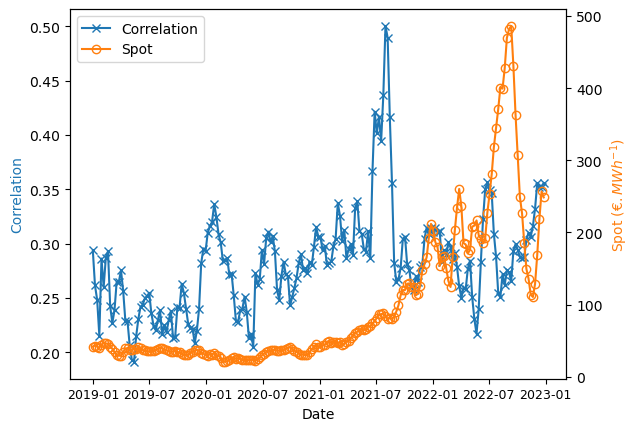} & \includegraphics[width=0.45\textwidth]{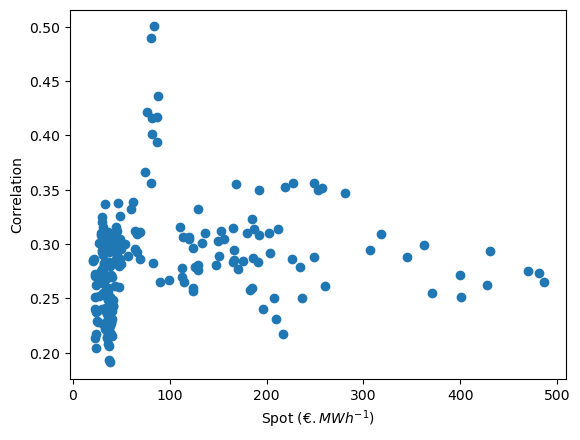}
    \end{tabular}
    \caption{Correlation proxy parameter \eqref{eq:correl_param} estimated each week from the last 28 trading sessions of the intraday market against the average spot price during the estimation period for Germany.}
    \label{fig:correl_ts}
\end{figure}

\begin{figure}[H]
    \centering
    \begin{tabular}{cc}
\includegraphics[width=0.45\textwidth]{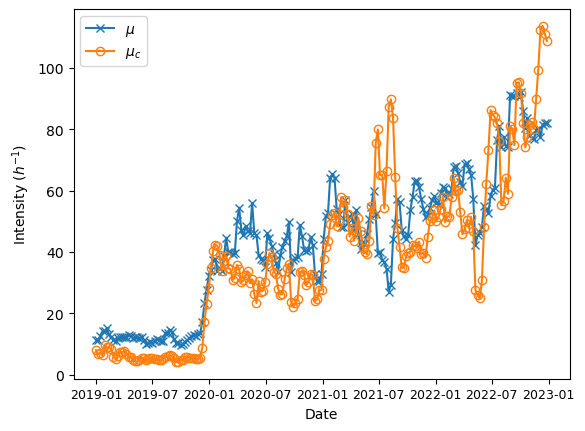} &
\includegraphics[width=0.45\textwidth]{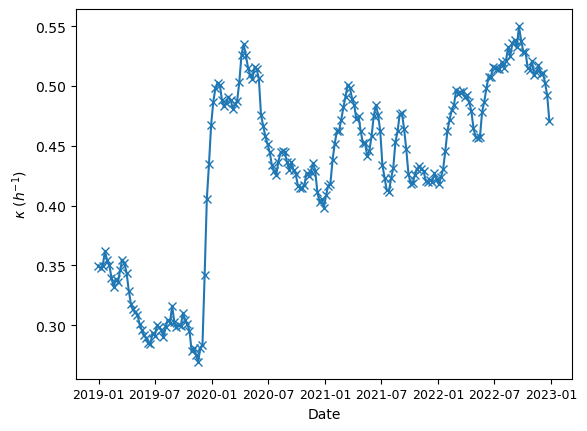}    
    \end{tabular}
    \caption{Time evolution of $\mu$ and $\mu_c$ (left) and $\kappa$ (right) estimated weekly from the last 28 trading sessions of the German intraday market.}
    \label{fig:other_parameters}
\end{figure}

\begin{figure}[H]
    \centering
    \begin{tabular}{cc}
    \includegraphics[width=0.45\textwidth]{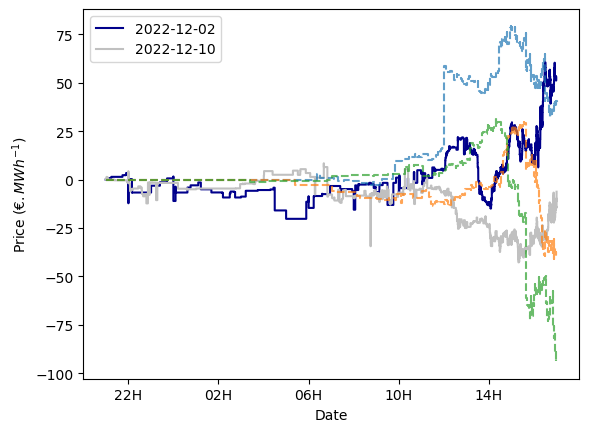} & \includegraphics[width=0.45\textwidth]{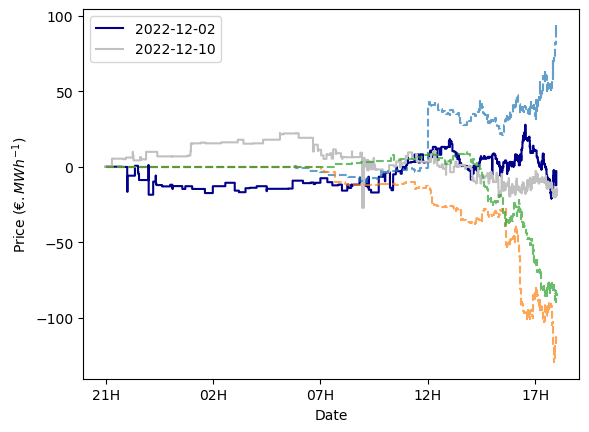}
    \end{tabular}
    \caption{Price simulations (dashed curves), starting from the same initial value with estimated parameters during the month of December 2022, together with two samples of data (solid curves), for maturities 18 (left) and 19 (right).}
    \label{fig:simu_model_ex}
\end{figure}

We now present the estimation results for France and Germany for the years 2019 to 2022. We consider the 24 hourly products ($M=24$), $\Delta = 30 $ minutes, $T_{b,m}=0$ and $T_{e,m} = T_m -1$ hour for $m=1,\ldots,M$ and $\delta = 1$ hour. In Figure~\ref{fig:reg_correl}, we show in solid green the correlation curve resulting from the estimated parameters of the model for Germany, together with the empirical correlations, as a function of the distance between maturities. The model manages to reasonably represent correlation levels and their decay. If we compare the solid green curve with the dashed orange curve, which is the least squares estimator between the empirical correlations and the function $x \to a\exp(-\frac{\kappa}{2}x)$, we get very similar results. In the model $\kappa$ is estimated on the evolution of the intensity over time with the least squares estimator \eqref{eq:empirical_ls_intensity}-\eqref{eq:sum_ls} and this value, estimated without taking into account the correlations between products, gives results relatively close to the direct estimation on the correlation curve (dashed orange curve). This justifies the use of a single $\kappa$ parameter to model both the Samuelson effect and the Samuelson correlation effect. Table~\ref{tab:params_de} and Table~\ref{tab:params_fr} show the estimated model parameters for each year for Germany and France respectively, as well as the first two moments of the jump sizes. The parameters $\mu$ and $\mu_c$ are higher for Germany than for France, which can be explained by a more liquid market in Germany, with more transactions. The parameter $\kappa$ is of the same order of magnitude from one country to another. The different parameters increase with the year, which is probably due to the increase in spot prices from 2019 to 2022. To support this assumption, we estimate the parameters every week (starting on Monday) from January 1\textsuperscript{st}, 2019 to December 31\textsuperscript{st}, 2022, using the last 28 days of data (i.e. 4 weeks). As previously done to remove outliers, for each period of estimation we remove returns that are larger in absolute value than 5 times the standard deviation of returns. Note that for each estimation period, the number of removed returns is less than $1\%$ of the total number of returns. The volatility parameter $\sigma$ defined in \eqref{eq:vol_param} is displayed in Figure~\ref{fig:vol_spot_de} against the average German spot price over the estimation period for German prices. 
There is a strong linear relationship between price volatility and spot prices. Future research could consider the spot price as a covariate in the intensity of our model. $\mu$ and $\mu_c$ evolve over time in the same way as shown in Figure~\ref{fig:other_parameters}: both seem to depend on the spot price. In contrast, the correlation parameter \eqref{eq:correl_param} does not seem to depend on the spot price, see Figure~\ref{fig:correl_ts}, which is consistent with a linear dependence of $\mu$ and $\mu_c$ on the spot price (as the correlation is a ratio). Finally, we show some simulation examples in Figure~\ref{fig:simu_model_ex}. There are some differences between the behaviour of the data and that of the simulations: 
\begin{itemize}
    \item[(i)] The prices from the simulations do not move at the beginning of the trading session, while the prices from the data show some jumps that seem to have mean-reverting behaviour.
    \item[(ii)] In general, the price from the data has more jumps than the one from the simulations, which is consistent with the estimation method: to avoid overestimating the volatility due to microstructural noise, we have considered the 30-minute return level. As volatility and intensity are closely related, this leads to an underestimation of the number of jumps, which in this case is related to round-trips on a microscopic scale.
    \item[(iii)] The proportion of big jumps at the end of the trading session in the simulations appears to be higher than in the data.
\end{itemize}
The following approaches could be considered to address these issues:
\begin{itemize}
\item[(i)] Consider a more complex time-dependent baseline with, for example, a first period consisting of a small and constant baseline and then a second period with an exponential function as in our model. 
\item[(ii)] Adding self-excitation to our model via Hawkes processes, which are relevant for marginal modelling of intraday prices, see~\cite{deschatre2023b}. This would also avoid the arbitrary choice of sampling time step $\Delta$ when estimating the covariance matrix.
\item[(iii)] Using a time-dependent law for the jumps, which is consistent with the results of Deschatre and Gruet~\cite{deschatre2023b} who have found that the distribution of jumps differs with time to maturity.
\end{itemize}

\begin{remark} Note the break in the intensity values at the beginning of 2020 for Germany: it may be due to the change in the frequency of the data (to the minute in 2019 and to the second in 2020), as there is more than one transaction per minute (if we look at $2(\mu+\mu_c)$ as a proxy at the beginning of 2020) and we have only taken into account the last transaction when several prices have the same timestamp. This phenomenon is not met in France as the intensity is much lower than one transaction per minute, the number of transactions in the same minute is much lower. The same phenomenon is observed for $\kappa$, see Figure~\ref{fig:other_parameters}, and can explain the increase in Table~\ref{tab:params_de} of $\kappa$ from 0.33 to 0.45.
\end{remark}

\begin{table}[H]
    \centering
\begin{tabular}{rrrrrr}
\toprule
 year &  $\kappa$ &  $\mu$ &  $\mu_c$ &  $\int_K y\nu(dy)$ &  $\int_K y^2\nu(dy)$ \\
\midrule
 2019 &      0.33 &  14.60 &     9.23 &               0.60 &                 0.35 \\
 2020 &      0.45 &  37.65 &    32.06 &               0.43 &                 0.18 \\
 2021 &      0.45 &  51.50 &    48.78 &               0.70 &                 0.49 \\
 2022 &      0.50 &  71.96 &    65.68 &               1.31 &                 1.72 \\
\bottomrule
\end{tabular}
\caption{Model parameters for German prices. The unit for $\kappa$, $\mu$ and $\mu_c$ is the inverse of the hour.}
\label{tab:params_de}
\end{table}

\begin{table}[H]
    \centering
\begin{tabular}{rrrrrr}
\toprule
 year &  $\kappa$ &  $\mu$ &  $\mu_c$ &  $\int_K y\nu(dy)$ &  $\int_K y^2\nu(dy)$ \\
\midrule
 2019 &      0.36 &   7.12 &     2.57 &               0.79 &                 0.62 \\
 2020 &      0.38 &  12.48 &     4.53 &               0.57 &                 0.33 \\
 2021 &      0.48 &  17.00 &     6.43 &               1.19 &                 1.42 \\
 2022 &      0.51 &  23.81 &    12.55 &               2.57 &                 6.62 \\
\bottomrule
\end{tabular}
    \caption{Model parameters for French prices. The unit for $\kappa$, $\mu$ and $\mu_c$ is the inverse of the hour.}
    \label{tab:params_fr}
\end{table}

\section{Battery valuation}
\label{sec:battery}

In the previous sections, we proposed a statistical model to simulate the different intraday prices over a trading session, taking into account their dependence structure. In order to assess the quality of the model, in addition to the statistical features that the model can represent, we consider the valuation of a storage and see the ability of our model to capture value in the intraday market through the volatility and correlation structure of the prices. Throughout this section, we will consider the case of a 2h battery and the case of a 3h battery.
A $n$h battery has the following characteristics:
\begin{itemize}
    \item[-] The battery capacity is $n$~MWh ;
    \item[-] the injection and withdrawal capacities are $1$ MWh per hour (i.e. a power of 1 MW).
\end{itemize}
Battery efficiency is taken to be equal to $\rho=0.92$, so:
\begin{itemize}
    \item[-] Injecting $E$ MWh into storage requires us to obtain $\frac{E}{\rho}$ MWh from the grid, and therefore to buy the corresponding quantity on the market;
    \item[-] withdrawing $E$ MWh from storage will only inject $\rho E$ MWh into the network, and therefore corresponds to the sale of this volume on the market.
\end{itemize}
We consider the intraday prices $f_{m,t}$ for delivery period $\left[T_m, T_m+\theta\right]$ with $\theta = 1$ hour, with $T_1=0,\ldots,T_{24}=23$ (each hour of the day). Each hour, a decision is taken on the nearest maturity, with a delay of $\Delta = $1 hour: the decision is therefore based on the price $f_{m, T_{m-1}}$ for each $m \in \{1,\ldots,24\}$.\\
The battery is managed assuming zero stock at $T_1=0$ hour each day.
The objective function is to maximise the expected gain at date $T_0 = 15$h on the day before management. The non-anticipatory control taken at date $T_i -\Delta$ belonging to $\mathcal{F}_{T_i-\Delta } = \{ f_{m, s} / s \le T_i -\Delta,\, m=1,\ldots,24\}$ is noted $C_i$ (a positive $C_i$ corresponds to an injection) and we note $\tilde C= (C_1, \dots, C_{24})$. \\
The function $J$ that we want to maximise is written as:
\begin{align}
\label{eq:optSto}
    J(\tilde C) =  -\E[  \sum_{i=1}^{24}  C_i ( \frac{1}{\rho} 1_{C_i \ge 0} + \rho 1_{C_i \le 0} ) f_{i,T_{i} - \Delta} / \mathcal{F}_{T_0}]
\end{align}
with the constraints for $i=1, \dots, 24$:
\begin{align*}
    0 \le & \sum_{j=1}^i C_j  \le \bar C, \\
    - \underline{C}&  \le C_i   \le \underline{C} .
\end{align*}
In our case, $\bar C= n$ and $\underline C= 1$.
To solve the problem \eqref{eq:optSto}, the classical technique consists in using dynamic programming in a Longstaff Schwartz approach \cite{warin2012gas}: at each explored stock level, generalising the results of Bardou et al.~\cite{bardou2009optimal} to our $n$ hours battery case, the optimal control is to be chosen in $\{ -\underline C, 0, \underline C\}.$
At each date, and for each stock level to be explored, we need to calculate conditional earnings expectations: knowing the future earnings trajectories at date $T_{i+1}$, $G_{i+1}$, we need to estimate  $A_i=\E[ G_{i+1} / {\mathcal F}_{T_{i}}]$.
 In this case, the available information is generated by the prices of future products for the rest of the day  
 
\begin{align}
  A_i = \E[ G_{i+1} / \{f_{j , T_i}\}_{j>i}]
  \label{eq:reg}
 \end{align}

 Unfortunately, calculating \eqref{eq:reg} is difficult because of the dimension of the problem.\\
 We therefore propose to approximate \eqref{eq:reg} by
 \begin{align}
 A_i \simeq A_i^p :=  \E[ G_{i+1} / \{f_{j , T_{i}}\}_{ i+p \ge j>i}]
 \end{align}
 
 for $p$ varying from $1$ to $6$.
 In the previous approximation, it is assumed that the information required is reduced to the observation of $p$ products of closest maturity. It is expected that increasing the number of factors will lead to an improvement in the result. This improvement is expected to diminish with increasing $p$.\\
 By limiting $p$ to $6$, classical techniques for calculating conditional expectation by regression on a basis of functions become usable.
 We choose to use local adaptive linear bases from \cite{bouchard2012monte} in the StOpt library \cite{gevret2018stochastic}.
 \begin{remark}
     There are alternative techniques to dynamic programming for dealing with the curse of dimension. These techniques, based on neural networks \cite{warin2023reservoir}, are effective, but the cost of their implementation does not allow them to be used for daily optimisation and backtesting, as in our article.
 \end{remark}
 To test our model, we perform the following backtest for each day of a year of a market for each level $p$ of approximation of the conditional expectation:
\begin{itemize}
     \item[-] A stochastic optimisation calculation is performed by solving \eqref{eq:optSto} with the parameters set for the market under study at the current day: we then recover the optimal strategy associated with intraday storage management for the model used. For this optimisation, we use $500000$ Monte Carlo price trajectories and $4$ meshes in each dimension for $p \le 4$, and one mesh  per dimension beyond 4, giving us a total of $4^{p \wedge 4}$ meshes.
    \item[-] We apply the previous strategy to intraday prices trajectory observed during the current day, giving us the value obtained in back test during the day.
\end{itemize}
The parameters are estimated every Monday from January, 1\textsuperscript{st}, 2019 to December 31\textsuperscript{st}, 2022, using the last 28 days of data as in Section~\ref{sec:numerical_results}.
The annual gain obtained by applying our strategy for a level $p$ of approximation of the conditional expectation is then deduced.
Our strategy can be compared to a ``spot control'' optimisation strategy: in this strategy,
the control is calculated from the spot prices $\{ f_{i,T_0} \}_{i=1,\ldots, 24}$ known  in $T_0$.
The optimal control $\tilde D= (D_1, \dots, D_{24})$ is obtained by maximizing the following problem:
\begin{align}
     \hat J( \tilde D) = - \sum_{i=1}^{24}  D_i ( \frac{1}{\rho} 1_{D_i \ge 0} + \rho 1_{D_i \le 0} )  f_{i,T_0}, \label{eq:optDet}
 \end{align}
under the constraints for  $i=1, \dots, 24$:
 \begin{align*}
     0 & \le \sum_{j=1}^i D_j  \le \bar C, \\
     - \underline{C} & \le D_i   \le \underline{C} .
 \end{align*}
  This control is applied in intraday, with the idea that the best view of intraday prices at $T_0$ is given by the spot price. We call this strategy the ``Spot'' strategy.
 In the Tables ~\ref{tab:Pinflue2}, ~\ref{tab:Pinflue3}, we evaluate the backtest influence of $p$ on the gains obtained by the Poisson model~\eqref{eq:model_pm}-\eqref{eq:model} for the year 2022 on the German market. As expected, an increase in $p$ improves the gains obtained. Gain is optimal for $p=4$. There is a slight deterioration for $p=5$  and $p=6$, which can be explained by the fact that the conditional expectation method would require more trajectories in optimisation and also that the number of days in backtesting is limited. Depending on the year, the market, the characteristics of the battery, the optimal choice for $p$ is $4$ or $5$. In the sequel we present results for $p=1,3$ and $5$ on the different cases.

\begin{table}[H]
     \centering
    \begin{tabular}{ccccccc}  \hline
       $p$ &  1 & 2 & 3 & 4 & 5 & 6 \\  \hline
        Gain  &  70728   & 102504   & 110642   & 112401  & 111725  & 109717    \\  
\bottomrule
     \end{tabular}
     \caption{Backtest value obtained on 2022 data as a function of $p$ for the Poisson model on the German market for a 2h battery.}
     \label{tab:Pinflue2}
 \end{table}

 \begin{table}[H]
     \centering
    \begin{tabular}{ccccccc}  \hline
       $p$ &  1 & 2 & 3 & 4 & 5 & 6 \\  \hline
        Gain  &  113921   & 142909   & 150434   & 153693  & 152955  & 152817    \\  
\bottomrule
     \end{tabular}
     \caption{Backtest value obtained on 2022 data as a function of $p$ for the Poisson model on the German market for a 3h battery.}
     \label{tab:Pinflue3}
 \end{table}

 In Table~\ref{tab:storage2H} and Table~\ref{tab:storage3H}, the sum of annual gains obtained by the spot strategy, the Poisson  model~\eqref{eq:model_pm}-\eqref{eq:model}, and the limit diffusion model \eqref{prop:trajectoryconv} are given as a function of market and year. The use of our model enables us to obtain better performances for the valuation of a battery compared to the ``Spot'' strategy. The ``Spot'' strategy sees only the price expectation and gains value from the shape of the prices. Our model allows us to add gains linked to the volatility and correlation structure of prices. Interestingly, the strategy derived from the diffusion limit model yields gains equivalent to those derived from the Poisson model. The approximation seems sufficient to solve the control problem in the simplified case of diffusion. The approximation of a control problem where the noise arises from a Poisson process 
 with a diffusive limit when the intensity becomes large, 
 is very recently studied theoretically by Abeille et al.~\cite{abeille2023}.

\smallskip
In this section, we have limited ourselves to a decision for the product with the closest maturity each hour. Our model makes it possible to optimise the battery by taking positions in several products at the same time, as all prices are simulated continuously over the trading session; however, liquidity costs would have to be modelled to make the problem realistic. We do not deal with this case here; it is an area for future research.

\begin{table}[H]
    \centering
\begin{minipage}[t]{0.49\linewidth}
\begin{tabular}{ccccc}\hline
      Year & p  & Spot     &  Poisson &  Diffusion  \\ 
\midrule
2019  &  1  &  14176  &  11648  &   11664   \\ 
2019  &  3  &  14176  &  21107  &   20126   \\ 
2019  &  5  &  14176  &  20972  &   21186   \\ 
\midrule
2020  &  1  &  16040  &  10799  &   10772   \\ 
2020  &  3  &  16040  &  21935  &   21821   \\ 
2020  &  5  &  16040  &  21982  &   21776   \\ 
\midrule
2021  &  1  &  35825  &  27301  &   27309   \\ 
2021  &  3  &  35825  &  45789  &   45387   \\ 
2021  &  5  &  35825  &  45810  &   45669   \\ 
\midrule
2022  &  1  &  88346  &  70728  &   70627   \\ 
2022  &  3  &  88346  &  110642  &   110311   \\ 
2022  &  5  &  88346  &  111725  &   111520   \\ 
\bottomrule
     \end{tabular}
 \caption*{German market.}
    \end{minipage}
\begin{minipage}[t]{0.49\linewidth}
\begin{tabular}{ccccc}\hline
      Year & p  & Spot     &  Poisson &  Diffusion  \\ 
\midrule
2019  &  1  &  14409  &  14252  &   14225   \\ 
2019  &  3  &  14409  &  16112  &   16034   \\ 
2019  &  5  &  14409  &  16164  &   16170   \\ 
\midrule
2020  &  1  &  14947  &  13579  &   13550   \\ 
2020  &  3  &  14947  &  17169  &   17130   \\ 
2020  &  5  &  14947  &  17510  &   17348   \\ 
\midrule
2021  &  1  &  38191  &  37595  &   37555   \\ 
2021  &  3  &  38191  &  42212  &   42180   \\ 
2021  &  5  &  38191  &  42177  &   42018   \\ 
\midrule
2022  &  1  &  81017  &  82956  &   82946   \\ 
2022  &  3  &  81017  &  97725  &   97835   \\ 
2022  &  5  &  81017  &  98346  &   98063   \\ 
\bottomrule
     \end{tabular}
 \caption*{French market.}
    \end{minipage}
\caption{Battery 2h }
   \label{tab:storage2H}
 \end{table}
\begin{table}[H]
    \centering
\begin{minipage}[t]{0.49\linewidth}
\begin{tabular}{ccccc}\hline
      Year & p  & Spot     &  Poisson &  Diffusion  \\ 
\midrule
2019  &  1  &  19781  &  19273  &   19270   \\ 
2019  &  3  &  19781  &  27920  &   27726   \\ 
2019  &  5  &  19781  &  28505  &   28605   \\ 
\midrule
2020  &  1  &  22105  &  19065  &   19053   \\ 
2020  &  3  &  22105  &  29668  &   29566   \\ 
2020  &  5  &  22105  &  29753  &   29626   \\ 
\midrule
2021  &  1  &  49113  &  44634  &   44525   \\ 
2021  &  3  &  49113  &  62070  &   61714   \\ 
2021  &  5  &  49113  &  63696  &   63314   \\ 
\midrule
2022  &  1  &  121030  &  113921  &   113852   \\ 
2022  &  3  &  121030  &  150434  &   150322   \\ 
2022  &  5  &  121030  &  152955  &   152870   \\ 
\bottomrule
     \end{tabular}
 \caption*{German market.}
    \end{minipage}
\begin{minipage}[t]{0.49\linewidth}
\begin{tabular}{ccccc}\hline
      Year & p  & Spot     &  Poisson &  Diffusion  \\ 
\midrule
2019  &  1  &  19436  &  19771  &   19778   \\ 
2019  &  3  &  19436  &  21636  &   21538   \\ 
2019  &  5  &  19436  &  21923  &   21799   \\ 
\midrule
2020  &  1  &  19982  &  19021  &   19015   \\ 
2020  &  3  &  19982  &  22628  &   22583   \\ 
2020  &  5  &  19982  &  23184  &   22820   \\ 
\midrule
2021  &  1  &  51069  &  51450  &   51466   \\ 
2021  &  3  &  51069  &  55143  &   55194   \\ 
2021  &  5  &  51069  &  55697  &   55560   \\ 
\midrule
2022  &  1  &  109236  &  114576  &   114559   \\ 
2022  &  3  &  109236  &  127632  &   127747   \\ 
2022  &  5  &  109236  &  129893  &   128570   \\ 
\bottomrule
     \end{tabular}
 \caption*{French market.}
    \end{minipage}
\caption{Battery 3h }
   \label{tab:storage3H}
 \end{table}

\section{Proofs}
\label{sec:proof}

\subsection{Proof of Proposition~\ref{prop:martingale}}
\label{sec:proofofmartingale}
Let \[
\overline{f}_{m,t}^{\perp,h} = \int_0^t \int_{\R_+} \int_K y {\bf 1}_{x \leq \mu e^{-\kappa(T_m-s)} {\bf 1}_{s \leq T_m}} \pi^{h}_m(ds,dx,dy) - \int_K y\nu(dy) \int_0^t \mu e^{-\kappa(T_m-s)} {\bf 1}_{s \leq T_m}ds
\]
and 
\[
\overline{f}_{m,t}^{c,h} = \int_0^t \int_{\R_+} \int_K y {\bf 1}_{x \leq \mu_c e^{-\kappa(T_m-s)} {\bf 1}_{s \leq T_m}} \pi^{h}_m(ds,dx,dy) - \int_K y\nu(dy) \int_0^t \mu_c e^{-\kappa(T_m-s)} {\bf 1}_{s \leq T_m}ds
\]
for $m=1,\ldots,M$ and $h=+,-$.
For $h=+,-$, the process 
\begin{equation} \label{eq:vector_martingale}
    \left(\overline{f}_{1,t}^{\perp,h},\overline{f}_{1,t}^{c,h},\ldots,\overline{f}_{M,t}^{\perp,h},\overline{f}_{M,t}^{c,h}\right)
\end{equation}
is a $(\mathcal{F}_t)_t$ squared integrable martingale. Hence, \\
$\left(f_{m,t}^{h} - \int_{\R} y \nu(dy) \int_0^t \left(\mu + \mu_c\right)e^{-\kappa(T_m-s)}{\bf 1}_{s \leq T_m}ds\right)_{m=1,\ldots,M}$ is also a squared integrable martingale for $h=+,-$ as a linear combination of the components of \eqref{eq:vector_martingale}.

\subsection{Proof of Proposition~\ref{prop:proba_common_jumps}}
\label{sec:proofofprobacommonjumps}

With $h \in \{+,-\}$ and $A_{m,u,t} = \{(s,x) | u\leq s \leq t,\, 0 \leq x \leq \mu_c e^{-\kappa(T_m - s)}\}$, we aim to compute
\[\mathbb{P}\left(\bigcap_{m \in \mathcal{M}_1}\{\pi^{h}\left(A_{m,u,t} \times K\right) \geq 1\} \cap \bigcap_{m \in \mathcal{M}_2}\{\pi^{h}\left(A_{m,u,t} \times K\right) = 0\}\right).\]
As $(A_{m,u,t})_{m=1,\ldots,M}$ is a decreasing sequence and $\pi^h$ is a counting measure,  
\[\bigcap_{m \in \mathcal{M}_1}\{\pi^{h}\left(A_{m,u,t} \times K\right) \geq 1\} = \{\pi^{h}\left(A_{\max \mathcal{M}_1,u,t} \times K\right) \geq 1\}\]
and
\[\bigcap_{m \in \mathcal{M}_2}\{\pi^{h}\left(A_{m,u,t} \times K\right) = 0\} = \{\pi^{h}\left(A_{\min \mathcal{M}_2,u,t} \times K\right) = 0\}.\]
If $\max \mathcal{M}_1 \geq \min \mathcal{M}_2$, then $A_{\max \mathcal{M}_1,u,t} \subset A_{\min \mathcal{M}_2,u,t}$ and
\[\{\pi^{h}\left(A_{\max \mathcal{M}_1,u,t} \times K\right) \geq 1\} \cap \{\pi^{h}\left(A_{\min \mathcal{M}_2,u,t} \times K\right) = 0\} = \emptyset.\]
Hence,
\[\mathbb{P}\left(\bigcap_{m \in \mathcal{M}_1}\{\pi^{h}\left(A_{m,u,t} \times K\right) \geq 1\} \cap \bigcap_{m \in \mathcal{M}_2}\{\pi^{h}\left(A_{m,u,t} \times K\right) = 0\}\right) = 0.\]
Otherwise, we have
\[\{\pi^{h}\left(A_{\max \mathcal{M}_1,u,t} \times K\right) \geq 1\} \cap \{\pi^{h}\left(A_{\min \mathcal{M}_2,u,t} \times K\right) = 0\}\]
equal to
\[\{\pi^{h}\left(\left(A_{\max \mathcal{M}_1,u,t} \backslash A_{\min \mathcal{M}_2,u,t}\right)\times K\right) \geq 1\} \cap \{\pi^{h}\left(A_{\min \mathcal{M}_2,u,t} \times K\right) = 0\}.\]
$\pi^{h}$ being a Poisson measure, and $\left(A_{\max \mathcal{M}_1,u,t} \backslash A_{\min \mathcal{M}_2,u,t}\right) \cap A_{\min \mathcal{M}_2,u,t} = \emptyset$, we get
\[\mathbb{P}\left(\{\pi^{h}\left(\left(A_{\max \mathcal{M}_1,u,t} \backslash A_{\min \mathcal{M}_2,u,t}\right)\times K\right) \geq 1\} \cap \{\pi^{h}\left(A_{\min \mathcal{M}_2,u,t} \times K\right) = 0\}\right)\]
equal to
\[\mathbb{P}\left(\{\pi^{h}\left(\left(A_{\max \mathcal{M}_1,u,t} \backslash A_{\min \mathcal{M}_2,u,t}\right)\times K\right) \geq 1\}\right)\mathbb{P}\left(\{\pi^{h}\left(A_{\min \mathcal{M}_2,u,t} \times K\right) = 0\}\right)\]
which is equal to
\[\left(1-\exp\left(-\int_{u}^t \mu_c \left(e^{-\kappa(T_{\max \mathcal{M}_1}-s)} - e^{-\kappa(T_{\min \mathcal{M}_2}-s)}\right)ds\right)\right) \exp\left(-\int_{u}^t \mu_c e^{-\kappa(T_{\min \mathcal{M}_2}-s)}ds\right).
\]

\subsection{Proof of Proposition~\ref{prop:simulation}}
\label{sec:proofofsimulation}

For $h=+,-$, $m=1,\ldots,M$, and $t \leq T$, we have

\[
\begin{split}
f^{c,h}_{m,t} =& \int_0^t {\bf 1}_{s \leq T_m}\int_{\R_+} \int_K y {\bf 1}_{0 \leq x \leq \mu_1(s)} \pi^{h}\left(ds, dx, dy\right) \\
&+ \sum_{i=2}^{m} \int_0^t {\bf 1}_{s \leq T_m}\int_{\R_+} \int_K y {\bf 1}_{\mu_{i-1}(s) < x \leq \mu_i(s)} \pi^{h}\left(ds, dx, dy\right)
\end{split}
\]

with $\mu_{i}(s) = \mu_c e^{-\kappa\left(T_m-s\right)}$ for $i=1,\ldots,M$, that is 
\begin{align*}
&f^{c,h}_{m} = \sum_{i=1}^{m} \int_K y \pi^{h}\left(A_{i,t}, dy\right), \,t\leq T_m,\\
&f^{c,h}_{m,t} = f^{c,h}_{m,T_m}, \,t\geq T_m,
\end{align*}
with $A_{1,t} = \{(s, x) | 0 \leq s \leq t, \, 0 \leq x \leq \mu_{1}(s)\}$ and $A_{i,t} = \{(s, x) | 0 \leq s \leq t, \, \mu_{i-1}(s) < x \leq \mu_{i}(s)\}$.
We get that $f^{c,h}_{m,t} = \int_0^t {\bf 1}_{s \leq T_m}\sum_{i=1}^m dP^{c,h}_{i,s}$ with $P^{c,h}_{i,t} = \int_K y \pi^{h}\left(A_{i,t}, dy\right)$, $i=1,\ldots,M$ and we show that $(P_{i,t}^{c,h})_t$ is an inhomogeneous compound Poisson process with intensity$\mu_1$ for $i=1$ and $\mu_i - \mu_{i-1}$ for  for $i=2,\ldots,M$ and jump law $\nu(dy)$. It remains to prove the independence of the $P_i^{c,h}$. Considering $t_1,\,t_2,\ldots,t_M$ $M$ real positive numbers, and $K_1,\,K_2,\ldots,K_M$ $M$ elements of $\mathcal{K}$ then the measures $\pi^{h}\left(A_{i,t_i}, K_i\right)$ for $i=1,\ldots,M$ are independent random variables since $\left(A_{i,t_i} \times K_i\right) \cap \left(A_{j,t_j} \times K_j\right) = \emptyset$ for $i \neq j$ as $0 < \mu_1 < \mu_2 < \ldots < \mu_M$ and $\pi^{h}$ is a Poisson measure. 

\subsection{Proof of Proposition~\ref{prop:moment1conv}}
\label{sec:proofofmoment1conv}
For $h=+,-$, the process
\[
n^{-1}f^{g,h,(n)}_t - \left(\int_{K}g(y)\nu(dy) \int_{0}^t \left(\tilde{\mu} + \tilde{\mu}_c\right)e^{-\kappa\left(T_m-s\right)} {\bf 1}_{s \leq T_m}ds\right)_{m=1,\ldots,M}
\]
is clearly a martingale (see the proof of Proposition~\ref{prop:martingale}) that we denote by $M^{(n)}$. Using the Burkholder-Davis-Gundy inequality, we get 
\[
\begin{split}
\mathbb{E}\left(\underset{t\in\left[0,T\right]}{\sup}\left\Vert  M^{(n)}_t\right\Vert^2\right) &\leq C \sum_{m=1}^M \mathbb{E}\left(\left[M_m^{(n)}, M_m^{(n)}\right]_T\right)\\
&=C n^{-2}\sum_{m=1}^M \mathbb{E}\left( \int_{0}^T \int_{\R_+} \int_{K}g^2(y){\bf 1}_{x \leq n\tilde{\mu} e^{-\kappa(T_m-s)}{\bf 1}_{s \leq T_m}} \pi_m^{h}(ds,dx,dy) +\right.\\
&\hspace{6em}\left.\int_{0}^T \int_{\R_+} \int_{K}g^2(y){\bf 1}_{x \leq n\tilde{\mu}_c  e^{-\kappa(T_m-s)}{\bf 1}_{s \leq T_m}} \pi^{h}(ds,dx,dy)\right)\\
&= Cn^{-1}\sum_{m=1}^M \int_{K}g(y)^2 \nu(dy) \int_0^T \left(\tilde{\mu} + \tilde{\mu}_c\right)e^{-\kappa(T_m-s)}{\bf 1}_{s \leq T_m}ds 
\end{split}
\]
and then $\mathbb{E}\left(\underset{t\in\left[0,T\right]}{\sup}\left\Vert  M^{(n)}_t\right\Vert^2\right)$ goes to 0 as $n \to \infty$, achieving the proof.

\subsection{Proof of Proposition~\ref{prop:trajectoryconv}}
\label{sec:proofoftrajectoryconv}

The quadratic covariation at time $t$ of the martingale $\frac{1}{\sqrt{n}}f^{n}$ is equal to 

\begin{equation} \label{eq:diag_part}
\begin{split}
&n^{-1}\int_0^t \int_{\R_+} \int_K y^2 {\bf 1}_{x \leq n\tilde{\mu}e^{-\kappa(T_m-s)} {\bf 1}_{s \leq T_m}} \pi_{m}^{+}(ds,dx,dy)\\
&+ n^{-1}\int_0^t \int_{\R_+} \int_K y^2 {\bf 1}_{x \leq n\tilde{\mu}_ce^{-\kappa(T_m-s)} {\bf 1}_{s \leq T_m}} \pi^+(ds,dx,dy)\\
&+n^{-1}\int_0^t \int_{\R_+} \int_K y^2 {\bf 1}_{x \leq n\tilde{\mu}e^{-\kappa(T_m-s)} {\bf 1}_{s \leq T_m}} \pi_{m}^{-}(ds,dx,dy)\\
&+ n^{-1}\int_0^t \int_{\R_+} \int_K y^2 {\bf 1}_{x \leq n\tilde{\mu}_ce^{-\kappa(T_m-s)} {\bf 1}_{s \leq T_m}} \pi^{-}(ds,dx,dy)
\end{split}
\end{equation}
on the diagonal at coordinate $m \in \{1,\ldots,M\}$ and 
\begin{equation}
    \label{eq:cov_matrix_nodiag}
n^{-1}\int_0^t \int_{\R_+} \int_K y^2 {\bf 1}_{x \leq n\tilde{\mu}_ce^{-\kappa(\max(T_k,T_l)-s)} {\bf 1}_{s \leq \min(T_k,T_l)}}\left(\pi^{+} + \pi^{-}\right)(ds,dx,dy)
\end{equation}
for $k,\,l \in \{1,\ldots,M\}$, $k\neq l$. From Proposition~\ref{prop:moment1conv}, the diagonal part \eqref{eq:diag_part} converges in $L^2(\mathbb{P})$ to 
\[2 \int_K y^2 \nu(dy) \int_0^t \left(\tilde{\mu} + \tilde{\mu}_c\right)e^{-\kappa(T_m-s)}{\bf 1}_{s \leq T_m} ds\]
as $n \to \infty$ and in the same way as the proof of Proposition~\ref{prop:moment1conv}, we prove that the remaining part of the quadratic covariation matrix \eqref{eq:cov_matrix_nodiag} converges to 
\[2 \int_K y^2 \nu(dy) \int_0^t  \tilde{\mu}_c e^{-\kappa(\max(T_k,T_l)-s)}{\bf 1}_{s \leq \min(T_k,T_l)} ds.\]
The limit of the covariation matrix corresponds to the covariation matrix of the process
\[
\left(\int_0^t \sqrt{\int_K y^2 \nu(dy)}\sqrt{2\left(\tilde{\mu} + \tilde{\mu}_c\right)} e^{-\frac{\kappa}{2}(T_m - s)}{\bf 1}_{s \leq T_m} dW_{m,s}\right)_{m=1,\ldots,M}
\]
with $W = \left(W_1, \ldots,W_M\right)^{\top}$ a multivariate Brownian motion with correlation matrix \\
$\left(\frac{\tilde{\mu} \delta_{kl} \tilde{\mu}_c}{\tilde{\mu} + \tilde{\mu}_c}e^{-\frac{\kappa}{2}|T_k-T_l|}   \right)_{k,l=1,\ldots,M}$. We achieve the proof by using \cite[Section VIII, Theorem 3.22]{jacod2013} which requires an additional Lindeberg-Feller type condition \cite[Section VIII, Theorem 3.22, Assumption 3.23]{jacod2013}, which we prove below.

\medskip
For $h=+,-$, the compensated measure of $\frac{1}{\sqrt{n}}\int_{x \in \R_+}{\bf 1}_{x \leq n\tilde{\mu} e^{-\kappa(T_m-s)}{\bf 1}_{s \leq T_m}}\pi_m^{h,(n)}(ds,dx,dy)$ is 
\[
n\tilde{\mu}e^{-\kappa(T_m-t)}{\bf 1}_{t \leq T_m}dt \otimes \nu(dy\sqrt{n})
\]
and for $\epsilon > 0$ and $t \geq 0$, using successively the Cauchy-Schwarz inequality and the Markov inequality, we get
\[
\begin{split} 
\int_{K} y^2 {\bf 1}_{y > \epsilon}\nu(dy\sqrt{n}) &\int_0^t n\tilde{\mu}e^{-\kappa(T_m-s)}{\bf 1}_{s\leq T_m}ds = \int_{K} y^2 {\bf 1}_{y > \sqrt{n}\epsilon}\nu(dy) \int_0^t \tilde{\mu}e^{-\kappa(T_m-s)}{\bf 1}_{s \leq T_m}ds\\
&\leq \sqrt{\int_{K} y^4\nu(dy)} \sqrt{\int_K{\bf 1}_{y > \sqrt{n}\epsilon}\nu(dy)} \int_0^t \tilde{\mu}e^{-\kappa(T_m-s)}{\bf 1}_{s \leq T_m}ds\\
&\leq n^{-1}\epsilon^{-2} \int_{K} y^4\nu(dy) \int_0^t \tilde{\mu}e^{-\kappa(T_m-s)}{\bf 1}_{s \leq T_m}ds.
\end{split}
\]
Therefore, 
\[\int_{K} y^2 {\bf 1}_{y > \epsilon}\nu(dy\sqrt{n}) \int_0^t n\tilde{\mu}e^{-\kappa(T_m-s)}{\bf 1}_{s \leq T_m}ds \to 0\]
as $n \to \infty$, which corresponds to \cite[Section VIII, Theorem 3.22, Assumption 3.23]{jacod2013}. This assumption is verified in the same way considering the compensated measure of \\ $\frac{1}{\sqrt{n}}\int_{x \in \R_+}{\bf 1}_{x \leq n\tilde{\mu}_c e^{-\kappa(T_m-s)}{\bf 1}_{s \leq T_m}}\pi^{h,(n)}(ds,dx,dy)$ for $h=+,-$.

\vip 
\section*{Acknowledgements}

The authors acknowledge support from the FiME Lab (Institut Europlace de Finance). The authors are grateful to F\'elix Trieu for his valuable help with the data and to Pierre Gruet for his constructive comments.

\bibliographystyle{plain}
\bibliography{biblio}

\begin{thebibliography}{10}

\bibitem{abeille2023}
Marc Abeille, Bruno Bouchard, and Lorenzo Croissant.
\newblock Diffusive limit approximation of pure-jump optimal stochastic control
  problems.
\newblock {\em Journal of Optimization Theory and Applications},
  196(1):147--176, 2023.

\bibitem{aid2022}
Ren{\'e} A{\"\i}d, Andrea Cosso, and Huy{\^e}n Pham.
\newblock Equilibrium price in intraday electricity markets.
\newblock {\em Mathematical Finance}, 32(2):517--554, 2022.

\bibitem{bacry2013a}
Emmanuel Bacry, Sylvain Delattre, Marc Hoffmann, and Jean-Fran{\c{c}}ois Muzy.
\newblock Modelling microstructure noise with mutually exciting point
  processes.
\newblock {\em Quantitative finance}, 13(1):65--77, 2013.

\bibitem{bacry2013b}
Emmanuel Bacry, Sylvain Delattre, Marc Hoffmann, and Jean-Fran{\c{c}}ois Muzy.
\newblock Some limit theorems for {H}awkes processes and application to
  financial statistics.
\newblock {\em Stochastic Processes and their Applications}, 123(7):2475--2499,
  2013.

\bibitem{balardy2022}
Clara Balardy.
\newblock An empirical analysis of the bid-ask spread in the continuous
  intraday trading of the german power market.
\newblock {\em The Energy Journal}, 43(3), 2022.

\bibitem{bardou2009optimal}
Olivier Bardou, Sandrine Bouthemy, and Gilles Pag{\`e}s.
\newblock Optimal quantization for the pricing of swing options.
\newblock {\em Applied Mathematical Finance}, 16(2):183--217, 2009.

\bibitem{barndorff2011}
Ole~E Barndorff-Nielsen.
\newblock Stationary infinitely divisible processes.
\newblock {\em Brazilian Journal of Probability and Statistics},
  25(3):294--322, 2011.

\bibitem{blasberg2019}
Alexander Blasberg, Nikolaus Graf~von Luckner, and R{\"u}diger Kiesel.
\newblock Modeling the serial structure of the {H}awkes process parameters for
  market order arrivals on the {G}erman intraday power market.
\newblock In {\em 2019 16\textsuperscript{th} International Conference on the
  {E}uropean {E}nergy {M}arket (EEM)}, pages 1--6. IEEE, 2019.

\bibitem{bouchard2012monte}
Bruno Bouchard and Xavier Warin.
\newblock Monte-{C}arlo valuation of {A}merican options: facts and new
  algorithms to improve existing methods.
\newblock In {\em Numerical Methods in Finance: Bordeaux, June 2010}, pages
  215--255. Springer, 2012.

\bibitem{bremaud2020}
Pierre Br{\'e}maud.
\newblock {\em Point process calculus in time and space}.
\newblock Springer, 2020.

\bibitem{chen2013}
Feng Chen and Peter Hall.
\newblock Inference for a nonstationary self-exciting point process with an
  application in ultra-high frequency financial data modeling.
\newblock {\em Journal of Applied Probability}, 50(4):1006--1024, 2013.

\bibitem{chen2016}
Feng Chen and Peter Hall.
\newblock Nonparametric estimation for self-exciting point processes-a
  parsimonious approach.
\newblock {\em Journal of Computational and Graphical Statistics},
  25(1):209--224, 2016.

\bibitem{tankov03}
Rama Cont and Peter Tankov.
\newblock {\em Financial modelling with jump processes}, volume~2.
\newblock CRC press, 2003.

\bibitem{deschatre2023}
Thomas Deschatre.
\newblock Adaptive estimation of intensity in a doubly stochastic {P}oisson
  process.
\newblock {\em Scandinavian Journal of Statistics}, 2023.

\bibitem{deschatre2021}
Thomas Deschatre, Olivier F{\'e}ron, and Pierre Gruet.
\newblock A survey of electricity spot and futures price models for risk
  management applications.
\newblock {\em Energy Economics}, 102:105504, 2021.

\bibitem{deschatre2023b}
Thomas Deschatre and Pierre Gruet.
\newblock Electricity intraday price modelling with marked {H}awkes processes.
\newblock {\em Applied Mathematical Finance}, pages 1--34, 2023.

\bibitem{favetto2019}
Benjamin Favetto.
\newblock The {E}uropean intraday electricity market: a modeling based on the
  {H}awkes process.
\newblock Available on hal.archives-ouvertes.fr, 2019.

\bibitem{finnah2022integrated}
Benedikt Finnah, Jochen G{\"o}nsch, and Florian Ziel.
\newblock Integrated day-ahead and intraday self-schedule bidding for energy
  storage systems using approximate dynamic programming.
\newblock {\em European Journal of Operational Research}, 301(2):726--746,
  2022.

\bibitem{fromont2011}
Magalie Fromont, B{\'e}atrice Laurent, and Patricia Reynaud-Bouret.
\newblock Adaptive tests of homogeneity for a {P}oisson process.
\newblock In {\em Annales de l'IHP Probabilit{\'e}s et statistiques},
  volume~47, pages 176--213, 2011.

\bibitem{gevret2018stochastic}
Hugo Gevret, Nicolas Langren{\'e}, Jerome Lelong, Xavier Warin, and Aditya
  Maheshwari.
\newblock {ST}ochastic {OPT}imization library in {C}++.
\newblock 2018.

\bibitem{graf2020}
Nikolaus Graf~von Luckner and R{\"u}diger Kiesel.
\newblock Modeling market order arrivals on the {G}erman intraday electricity
  market with the {H}awkes process.
\newblock {\em Journal of Risk and Financial Management}, 14(4), 2021.

\bibitem{hirsch2022}
Simon Hirsch and Florian Ziel.
\newblock Simulation-based forecasting for intraday power markets: Modelling
  fundamental drivers for location, shape and scale of the price distribution.
\newblock {\em arXiv preprint arXiv:2211.13002}, 2022.

\bibitem{hirsch2023}
Simon Hirsch and Florian Ziel.
\newblock Multivariate simulation-based forecasting for intraday power markets:
  Modelling cross-product price effects.
\newblock {\em arXiv preprint arXiv:2306.13419}, 2023.

\bibitem{jacod2013}
Jean Jacod and Albert Shiryaev.
\newblock {\em Limit theorems for stochastic processes}, volume 288.
\newblock Springer Science \& Business Media, 2013.

\bibitem{jaeck2016}
Edouard Jaeck and Delphine Lautier.
\newblock Volatility in electricity derivative markets: The {S}amuelson effect
  revisited.
\newblock {\em Energy Economics}, 59:300--313, 2016.

\bibitem{kiesel2017}
R{\"u}diger Kiesel and Florentina Paraschiv.
\newblock Econometric analysis of 15-minute intraday electricity prices.
\newblock {\em Energy Economics}, 64:77--90, 2017.

\bibitem{kramer2021}
Anke Kramer and R{\"u}diger Kiesel.
\newblock Exogenous factors for order arrivals on the intraday electricity
  market.
\newblock {\em Energy Economics}, 97:105186, 2021.

\bibitem{kremer2021}
Marcel Kremer, R{\"u}diger Kiesel, and Florentina Paraschiv.
\newblock An econometric model for intraday electricity trading.
\newblock {\em Philosophical Transactions of the Royal Society A},
  379(2202):20190624, 2021.

\bibitem{lindskog2003}
Filip Lindskog and Alexander~J McNeil.
\newblock Common {P}oisson shock models: applications to insurance and credit
  risk modelling.
\newblock {\em ASTIN Bulletin: The Journal of the IAA}, 33(2):209--238, 2003.

\bibitem{longstaff2001valuing}
Francis~A Longstaff and Eduardo~S Schwartz.
\newblock Valuing {A}merican options by simulation: a simple least-squares
  approach.
\newblock {\em The review of financial studies}, 14(1):113--147, 2001.

\bibitem{machlev2020}
R~Machlev, N~Zargari, NR~Chowdhury, J~Belikov, and Y~Levron.
\newblock A review of optimal control methods for energy storage systems-energy
  trading, energy balancing and electric vehicles.
\newblock {\em Journal of Energy Storage}, 32:101787, 2020.

\bibitem{narajewski2020}
Micha{\l} Narajewski and Florian Ziel.
\newblock Ensemble forecasting for intraday electricity prices: Simulating
  trajectories.
\newblock {\em Applied Energy}, 279:115801, 2020.

\bibitem{powojowski2002}
Miro~R Powojowski, Diane Reynolds, and Hans~JH Tuenter.
\newblock Dependent events and operational risk.
\newblock {\em Algo Research Quarterly}, 5(2):65--73, 2002.

\bibitem{reynaud2003}
Patricia Reynaud-Bouret.
\newblock Adaptive estimation of the intensity of inhomogeneous {P}oisson
  processes via concentration inequalities.
\newblock {\em Probability Theory \& Related Fields}, 126(1), 2003.

\bibitem{schneider2018}
Lorenz Schneider and Bertrand Tavin.
\newblock From the {S}amuelson volatility effect to a {S}amuelson correlation
  effect: An analysis of crude oil calendar spread options.
\newblock {\em Journal of Banking \& Finance}, 95:185--202, 2018.

\bibitem{veraart2023}
Almut~ED Veraart.
\newblock Periodic trawl processes: Simulation, statistical inference and
  applications in energy markets.
\newblock {\em arXiv preprint arXiv:2303.04121}, 2023.

\bibitem{warin2012gas}
Xavier Warin.
\newblock Gas storage hedging.
\newblock In {\em Numerical methods in finance: Bordeaux, June 2010}, pages
  421--445. Springer, 2012.

\bibitem{warin2023reservoir}
Xavier Warin.
\newblock Reservoir optimization and machine learning methods.
\newblock {\em EURO Journal on Computational Optimization}, page 100068, 2023.

\end{thebibliography}

\end{document}